\documentclass[fleqn,usenatbib]{mnras}
\usepackage{amsmath}
\usepackage{graphicx}
\usepackage{hyperref}
\usepackage{ragged2e}
\usepackage{comment}

\newcommand{\yangedit}[1]{\textcolor{black}{#1}}
\newcommand{\revision}[1]{\textcolor{black}{#1}}
\newcommand{\revisiontwo}[1]{\textcolor{black}{#1}}

\title{Ion densities of cold clouds driven by galactic outflows}
\author[L. Yang et al.]{
Lisiyuan Yang,$^{1}$\thanks{E-mail: lyang@umass.edu}
Neal Katz,$^{1}$
Evan Scannapieco$^{2}$
and Marcus Br\"uggen,$^{3}$
\\
$^{1}$Astronomy Department, University of Massachusetts, Amherst, MA 01003, USA\\
$^{2}$School of Earth and Space Exploration, Arizona State University, P.O. Box 871404, Tempe, AZ 85287-1404, USA\\
$^{3}$Universist\"at Hamburg, Hamburger Sternwarte, Gojenbergsweg 112, D-21029, Hamburg, Germany
}

\begin{document}
\maketitle
\begin{abstract}
	Observations of the circumgalactic medium (CGM) often display coincident absorption from species with widely varying ionization states, providing direct evidence for complex, multiphase interactions. Motivated by these measurements, we perform a series of cloud-crushing simulations that model cold clouds traveling through the hot CGM. We analyze the ion distributions of these clouds \revision{under the assumption of ionization equilibrium in the presence of a photoionizing background}, generate mock absorption spectra, and study their implications on quasar (QSO) absorption observations. Our results show multiphase features, in which ions with significantly different ionization potentials exist in the same absorber and share similar spectral features. However, our simulations are unable to explain high ions like O \textsc{vi} and their coexistence with lower ions as observed in many QSO absorption systems.
\end{abstract}

\section{Introduction}
The circumgalactic medium (CGM) and intergalactic medium (IGM) play an important role in the evolution of galaxies \citep{Putman2012, Tumlinson2017}. They are perturbed and enriched in metals by galactic outflows \citep{Tumlinson2011} from both stellar \citep{Heckman1990, Tremonti2004, Tripp2011, Chisholm2016, Johnson2017} and AGN feedback \citep{Scannapieco2004, DiMatteo2005, Fabian2012, King2015}, and hence serve as probes of these processes. Properties of the CGM have also been found to correlate with the properties of host halos, providing constraints on cosmological structure formation
\citep{Chen2010, Churchill2013, Werk2014, Tumlinson2017, Bordoloi2018}.

There are several ways to probe the CGM and IGM. X-ray emission through thermal bremsstrahlung (e.g. \citealt{Wang2014, Anderson2015}) and the Sunyaev-Zel'dovich (SZ) effect (e.g. \citealt{Birkinshaw1999, Lim2020}) can be used to detect the hot, virialized phase, especially around massive halos and clusters. In the presence of a background QSO, or if the host galaxy itself has a strong UV continuum, observations of absorption lines of H \textsc{i} and metal ions probe the cooler, photoionized components of the CGM and IGM (e.g. \citealt{Bechtold2001, Tripp2008, Danforth2008, Stocke2013, Danforth2016, Chisholm2018}). Compared to X-ray emission and the SZ effect, such absorption line studies are restricted by the requirement of a continuous UV background source. Furthermore, QSO absorption lines are necessarily restricted to pencil beams through the halo gas. Despite these limits, they have the unique advantage of being more sensitive to low column densities and have the ability to probe different metal ions with a wide range of ionization potentials.

Through these observations, the CGM has been found to be multiphase and complex in nature. Numerous studies have discovered that a large number of QSO absorption systems likely consist of two or more ionization phases (e.g. \citealt{Tripp2008, Meiring2013, Haislmaier2020, Zahedy2021}). Some samples even show ions with vastly different ionization potentials that are kinematically well-aligned and have similar line widths, indicating a physical connection between the various phases.

On a larger scale, statistical studies of the Ly$\alpha$ forest \citep{Oort1981} from the CGM and IGM can be used to measure cosmological perturbations of baryons and the evolution of the Universe \citep{Hernquist1996, Weinberg2003, Seljak2006, Kirkman2007, Baur2016}, and thereby provide constraints on cosmic models.

On the simulation side, the CGM and IGM are also of great interest. Since they are affected by gas accretion and feedback processes, simulated CGM and IGM ion distributions are frequently compared with QSO absorption observations to study inflows and outflows \citep{Oppenheimer2006, Dave2010, Ford2013, Hummels2013, Ford2014, Ford2016, Oppenheimer2018, Kauffmann2019, Appleby2021}. These studies have reinforced the multiphase nature of the CGM and stressed the importance of inflows and outflows.

However, it has been shown through zoom-in simulations that the scale lengths of local density and pressure variations in the CGM are at least as small as several hundred parsecs (e.g. \citealt{Lochhaas2023}), which is close to the typical sizes of single absorption clouds inferred from QSO absorption observations \citep{Rudie2019, Haislmaier2020}. Tracing the evolution of these cool clouds embedded in the hot phase requires even higher resolutions (e.g. \citealt{McCourt2018}). This is far out of reach of any cosmological-sized or even zoom-in simulation and poses a significant challenge to our theoretical understanding of QSO absorption-line systems, and of galaxy formation itself. 

To bridge the gap, numerous small-scale simulations have been conducted that focus on metal-rich clouds in galactic outflows traveling supersonically in a hot ambient medium (e.g. \citealt{Klein1994, Scannapieco2015, Brueggen2016, Schneider2017, BandaBarragan2019, BandaBarragan2020, BandaBarragan2021, Casavecchia2024}). Through such studies, a wide variety of physical processes, including radiative cooling, thermal conduction, magnetic fields, and chemical inequilibrium, are found to have significant impacts on the evolution of the clouds. 

Studies of the ion distributions and absorption spectra of these simulations are relatively few (e.g. \citealt{Cottle2018, Cruz2021, Casavecchia2024}), and most of them only focus on down-the-barrel observations, i.e. lines of sight along the direction that the clouds travel. In addition, theoretical works use various simplifying assumptions (for example, \citealt{Cottle2018} used the mean column densities and broadenings of entire projections to generate single Gaussian profiles, while \citealt{Cruz2021} ignored thermal broadening in their synthetic lines).

To implement some of the findings of these cloud crushing simulations into cosmological simulations, we have developed the PhEW (Physically Evolved Winds) model \citep{Huang2020a, Huang2021}, an analytical subgrid model for cosmological hydrodynamic simulations that tracks galactic outflows from stellar feedback. In cosmological simulations that use this model, after a wind particle is ejected by stellar feedback and reaches the CGM, it is converted into a PhEW particle. PhEW particles are modeled as collections of cool clouds that travel supersonically through the hot CGM, with an initial temperature of $10^4\ \mathrm{K}$ and a fixed initial cloud mass of $10^4$ to $10^5 M_\odot$. Their evolution is analytically tracked using results from the cloud crushing simulations conducted by \citet{Scannapieco2015} and \citet{Brueggen2016}, until they fully dissolve into the hot ambient medium. \citet{Huang2021} have shown that this model is robust against changes in cloud parameters, choice of hydrodynamic method, and numerical resolution, unlike other approaches. They also find that the PhEW model enhances wind recycling and metal mixing from the outflows into the hot phase of the CGM and IGM.

In this work, we carry out a series of cloud-crushing simulations similar to those performed by \citet{Scannapieco2015} and \citet{Brueggen2016}, but with a lower cloud density that matches the CGM of a typical Milky Way-like galaxy. Using these simulations, we compute the ion distributions of the clouds, generate mock absorption spectra, and subsequently fit them to study the statistical properties of individual, distinct absorbers, and how they are affected by the simulation parameters. We compare our results with observations and discuss the possible implications of our results on some of the commonly observed absorption features. Our results can also be applied to cosmological simulations that use physics-driven, multiphase subgrid wind models, such as the PhEW model, to study the effect of stellar feedback on quasar absorption (Yang et al., in prep).

This paper is organized as follows. In \S \ref{sec:methods}, we introduce our cloud-crushing simulations and the physical parameters characterizing our clouds. In \S \ref{sec:mass_loss} we track the mass evolution of the clouds in our simulations and discuss the physical factors that affect them. In \S \ref{sec:ion_densities} we describe our method for computing the ion distributions, generating mock spectra and fitting them to obtain lists of individual absorbers in our simulations. We also display examples of the ion maps and mock spectra. In \S \ref{sec:column_densities} we analyze the statistical distributions of the column densities and line widths of the absorbers in our simulations, explore their dependence on multiple physical parameters, and study the correlation between different ions. Our conclusions are presented in \S \ref{sec:conclusions}.
 
\section{Methods}\label{sec:methods}
\subsection{Simulations}\label{subsec:simulations}
We perform a set of cloud-crushing simulations similar to those carried out by \citet{Scannapieco2015} and \citet{Brueggen2016}. In our simulations, a \revision{uniform} spherical cloud of mass $M_{\mathrm{c}}=10^5\ M_{\odot}$, density $\rho_{\mathrm{c}}=10^{-26}\ \mathrm{g/cm^3}$ , radius $R_{\mathrm{c}}=(3M_{\mathrm{c}}/4\pi \rho_{\mathrm{c}})^{1/3}=545\ \mathrm{pc}$  \revisiontwo{and temperature $T_{\mathrm{c}}=10^4\ \mathrm{K}$} is placed in a supersonic hot flow that is in thermal pressure equilibrium with the cloud. Our clouds have a lower density than the ones used in \citet{Scannapieco2015} and \citet{Brueggen2016}, which have an initial density of $\rho_{\mathrm{c,old}}=10^{-24}\ \mathrm{g/cm^3}$, as they modeled down-the-barrel observations, and hence their clouds were expected to be much closer to the central galaxies. Our initial conditions of the clouds are made to match the pressure of the CGM of a Milky Way-like galaxy between redshifts 0 and 2 (see Appendix \ref{app:cloud_density}).

The system is evolved using the grid-based FLASH code \citep{Fryxell2000}, including radiative cooling, and either including or excluding thermal conduction. The equations governing the evolution of the system are identical to those given in \citet{Brueggen2016}:
\begin{align}
	\partial_t \rho+\nabla \cdot(\rho \boldsymbol{u}) &= 0, \label{eq:cont}\\
	\rho\left[\partial_t \boldsymbol{u}+(\boldsymbol{u} \cdot \nabla) \boldsymbol{u}\right] &= -\nabla p, \label{eq:momentum}\\
	\partial_t E+\nabla \cdot[E \boldsymbol{u}]  = -\nabla \cdot(p \boldsymbol{u}) &- n^2 \Lambda(T)+\nabla \cdot \boldsymbol{q}, \label{eq:energy}
\end{align}
where $\rho$ is the density, $\boldsymbol{u}$ is the velocity, $p=k_{\mathrm{B}}T\rho/(\mu m_{\mathrm{p}})$ is the pressure and $E=p /(\gamma-1)+\frac{1}{2} \rho|\boldsymbol{u}|^2$ is the energy density. \revision{The thermal conduction rate, $q$, is 0 in the non-conductive runs and is given by}
\begin{equation}
	\boldsymbol{q}=\min \left\{\begin{array}{l}
		\kappa(T) \nabla T \\
		0.34 n_e k_{\mathrm{B}} T c_{\mathrm{s}, \mathrm{e}} \nabla T/|\nabla T|,
	\end{array}\right.
\end{equation}
\revision{in the conductive runs}. Here
\begin{equation}
	\kappa(T)=\eta\cdot 5.6\times 10^{-7}T^{5/2} \mathrm{erg\ s^{-1}\ K^{-1}\ cm^{-1}},
	\label{eq:kappa}
\end{equation}
and $c_{\mathrm{s},e}=(k_{\mathrm{B}}T/m_{\mathrm{e}})^{1/2}$. 

We use a factor $\eta$ to model the suppression of thermal conduction by magnetic fields: $\eta=1$ for full Spitzer conduction and $0.1$ for weak conduction. \yangedit{Only isotropic conduction is considered in this work, and our simulations do not model magnetic fields directly. We also do not track the ion populations in a time-dependent, i. e., non-equilibrium ionization manner. For the cooling rate $\Lambda$, we use tables compiled by \citet{Wiersma2009} from the Cloudy code \citep{Ferland1998}}\revisiontwo{, assuming the default solar metal abundances used by Cloudy, which are tabulated in Table 1 of \citet{Wiersma2009}.}  Since the radii of our clouds are larger than those presented in \citet{Scannapieco2015} and \citet{Brueggen2016}, by a factor of $R_{\mathrm{c}}/R_{\mathrm{c,old}}=5.45$, we increase the size of the simulation box and the spatial resolution \revision{accordingly, to a volume of $8.7\ \mathrm{kpc}\times 8.7\ \mathrm{kpc}$ in the $x$ and $z$ direction and $6.5\ \mathrm{kpc}$ in the $y$ direction. The finest resolution is likewise decreased to $8.5\ \mathrm{pc}$.} Apart from these changes, our simulations are identical to theirs. 

The simulations are characterized by two dimensionless numbers: the Mach number of the hot flow $M_{\mathrm{hot}}$ and the initial density contrast $\chi_0=\rho_{\mathrm{c}}/\rho_{\mathrm{hot}}$, in addition to the level of thermal conduction. Table \ref{tab:run_list} lists the cloud crushing simulations that are used in this work, where $t_{\mathrm{cc}}$, the cloud crushing time, is given by \citep{Klein1994}
\begin{equation}
	t_{\mathrm{cc}} = \frac{\chi_0^{1/2} R_{\mathrm{c}}}{v_{\mathrm{hot}}},
\end{equation}
and $t_{\mathrm{cool}}$, the cooling time, is given by equation 7 of \citet{Scannapieco2015}. Most of our simulations have identical values of $M_{\mathrm{hot}}$ and $\chi_0$ to their counterparts in \citet{Scannapieco2015} and  \citet{Brueggen2016}\revision{, with ambient temperatures ranging from $3\times 10^6\ \mathrm{K}$ to $3\times 10^7\ \mathrm{K}$ and velocities from $1000\ \mathrm{km/s}$ to $3000\ \mathrm{km/s}$}. \revision{As discussed in \citet{Scannapieco2015}, \revisiontwo{where they used the \citet{Chevalier1985} model to calculate the Mach numbers and velocities of spherical outflows at different radii theoretically,} these parameters cover a wide range of outflow scenarios.} The only exceptions are the T0.1\_v150\_chi100 and T0.1\_v150\_chi100\_cond\_0.1 simulations, which model clouds with low Mach numbers and density contrasts that appear in some cosmological simulations, including those that use the PhEW model (Yang et al., in prep).

In this work, we mainly focus on the non-conductive and weakly-conductive simulations, while using the fully-conductive simulations occasionally as comparisons to the other conduction levels. As discussed in \S \ref{sec:mass_loss}, the clouds in the fully-conductive simulations tend to evaporate much faster than their non-conductive and weakly-conductive counterparts.

\begin{table*}
    \centering
    \caption{Cloud-crushing simulations used in this work.}
    \begin{tabular}{l c c c c c c c}
    \hline
    Name & Conduction$^{a}$ & $M_{\mathrm{hot}}$$^{b}$ & $v_{\mathrm{hot}}\ (\mathrm{km/s})$ & $T_{\mathrm{hot}}\ (10^6 \mathrm{K})$ & $\chi_0$ & $t_{\mathrm{cc}}\ (\mathrm{Myr})$ & $t_{\mathrm{cool}}\ (\mathrm{Myr})$ \\\hline
    T0.3\_v1000\_chi300 & no & 3.8 & 1000 & 3 & 300 & 9.17 & 0.17 \\
    T0.3\_v1000\_chi300\_cond\_0.1 & weak & 3.8 & 1000 & 3 & 300 & 9.17 & 0.17 \\
    T0.3\_v1000\_chi300\_cond & full & 3.8 & 1000 & 3 & 300 & 9.17 & 0.17 \\
    T0.3\_v1700\_chi300 & no & 6.5 & 1700 & 3 & 300 & 5.40 & 0.38 \\
    T0.3\_v1700\_chi300\_cond\_0.1 & weak & 6.5 & 1700 & 3 & 300 & 5.40 & 0.38 \\
    T0.3\_v1700\_chi300\_cond & full & 6.5 & 1700 & 3 & 300 & 5.40 & 0.38 \\
    T0.3\_v3000\_chi300 & no & 11.4 & 3000 & 3 & 300 & 3.06 & 3.56 \\
    T0.3\_v3000\_chi300\_cond\_0.1 & weak & 11.4 & 3000 & 3 & 300 & 3.06 & 3.56 \\
    T0.3\_v3000\_chi300\_cond & full & 11.4 & 3000 & 3 & 300 & 3.06 & 3.56 \\
    T1\_v1700\_chi1000 & no & 3.5 & 1700 & 10 & 1000 & 9.85 & 0.17 \\
    T1\_v1700\_chi1000\_cond\_0.1 & weak & 3.5 & 1700 & 10 & 1000 & 9.85 & 0.17 \\
    T1\_v1700\_chi1000\_cond & full & 3.5 & 1700 & 10 & 1000 & 9.85 & 0.17 \\
    T3\_v3000\_chi3000 & no & 3.6 & 3000 & 30 & 3000 & 9.67 & 0.18 \\
    T3\_v3000\_chi3000\_cond\_0.1 & weak & 3.6 & 3000 & 30 & 3000 & 9.67 & 0.18 \\
    T3\_v3000\_chi3000\_cond & full & 3.6 & 3000 & 30 & 3000 & 9.67 & 0.18 \\
    T0.1\_v150\_chi100 & no & 1.0 & 150 & 1 & 100 & 35.3 & 0.76 \\
    T0.1\_v150\_chi100\_cond\_0.1 & weak & 1.0 & 150 & 1 & 100 & 35.3 & 0.76 \\\hline
    \end{tabular}
    
    \raggedright
    $^{a}$Here ``weak" means $\eta=0.1$ in Equation \ref{eq:kappa}.\\
    \revision{$^{b}$Here $M_{\mathrm{hot}}$ refers to the Mach number of the ambient gas.}
    \label{tab:run_list}
\end{table*}

\subsection{Cloud selection}\label{subsec:cloud_cut}
The goal of our study is to explore the ion distributions due to the presence of the cloud. Therefore, it is crucial to separate the cloud region from its ambient material. Since FLASH is a grid-based code, a criterion must be employed to identify grid cells that belong to the cloud region. In \citet{Scannapieco2015}, to determine the mass of the cloud after partial disruption, they tested three different density criteria: $1$, $1/3$, and $1/10$ times the initial cloud density, where grid cells with densities above these criteria were attributed to the cloud. They found that changing the density criteria did not significantly affect the cloud mass, and ultimately settled for a criterion $\rho > \rho_{\mathrm{c}}/3$, which was subsequently used by \cite{Cottle2018} in their absorption line study. However, when considering the distribution of ion densities within the cloud, special caution must be exercised. When changing the criterion, most of the regions affected are on the edge of the cloud. These regions are usually small in volume and have relatively low densities, but their temperatures are generally higher than the central part of the cloud owing to interactions with the ambient gas. Therefore, even though they do not make up a considerable portion of the cloud mass, they can potentially contribute significantly to middle and high ion absorption.

To account for this, we adopt a different criterion for the cloud region that uses the tracer mass fraction $C_{\text{cloud}}$ \citep{Xu1995, Orlando2005}. This is a scalar field implemented in our simulations that equals the fraction of mass within each grid cell that originates from the cloud region in the initial setup. We define the cloud region as the collection of grid cells that are dominated by cloud material, i.e. those that have
\begin{equation}
	C_{\text{cloud}} \geq 0.5.
	\label{eq:cloud_cut}
\end{equation}
We find that this criterion accounts for the fragments of the clouds much better than density-based criteria. If not specified, ``cloud region" refers to regions that satisfy criterion (\ref{eq:cloud_cut}) hereafter.

\section{Evolution of the clouds}
\label{sec:mass_loss}
\citet{Scannapieco2015} and \citet{Brueggen2016} primarily focused on 4 specific snapshots from each simulation, namely those with remaining cloud masses roughly equal to 90\%, 75\%, 50\%, and 25\% of their initial mass. In our work, we also focus on snapshot times that have remaining cloud mass fractions close to these percentages, which we label t90, t75, t50, and t25. It should be noted, however, that our cloud masses are based on cloud criterion (\ref{eq:cloud_cut}) rather than the density criterion used by \citet{Scannapieco2015} and \citet{Brueggen2016}. Two processes cause mass loss from the clouds: Kelvin-Helmholtz instabilities (KHI), and, in cases with thermal conduction, conductive evaporation.

\begin{figure*}
	\centering
	\includegraphics[width=0.9\linewidth]{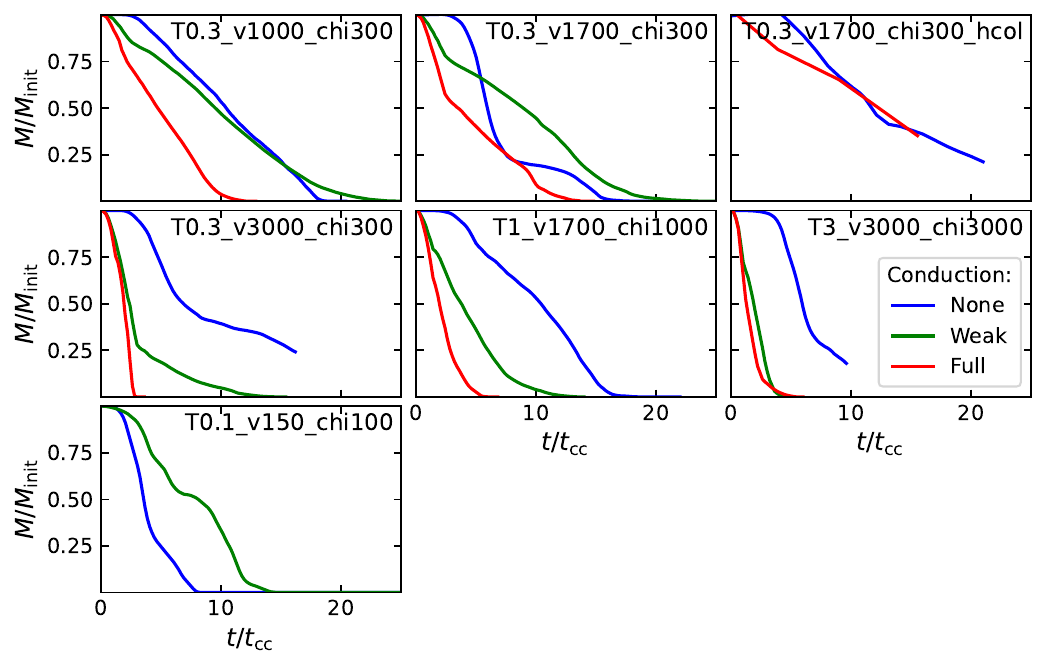}
	\caption{Mass loss curves of the clouds in our simulations. \revisiontwo{Here, names such as ``T0.3\_v1000\_chi300" refers to collections of 3 (sometimes 2) simulations with different levels of thermal conduction, which share the same ambient temperatures and velocities.} The colours are indicative of the conduction levels. As a comparison, two simulations from \citet{Scannapieco2015} and \citet{Brueggen2016} (dubbed ``hcol'') are also shown.}
	\label{fig:mass_loss}
\end{figure*}
In Figure \ref{fig:mass_loss}, we show the mass loss curves of all the simulations tabulated in Table \ref{tab:run_list}. We use different colours to depict simulations with different levels of thermal conduction.  As demonstrated in \citet{Brueggen2016}, thermal conduction has two competing effects on the survival of the cloud. On the one hand, it suppresses the KHI, thereby prolonging the survival time. On the other hand, thermal conduction causes evaporation from the surface of the cloud, which can shorten its lifetime. 

It is clear that in our simulations, conductive evaporation is very efficient at full Spitzer conduction and reduces the lifetime of the clouds in all cases. The most extreme case of this is found in the T0.3\_v3000\_chi300\_cond simulation, where the cloud blows up early on, leaving only a diffuse shell of evaporated gas. In contrast, weak conduction sometimes lowers the mass loss rates of the clouds by suppressing the KHI but not being strong enough to cause rapid evaporation. \yangedit{Morphologically, the evaporative pressure compresses the clouds in the conductive simulations, making them more compact than their non-conductive counterparts.}
\begin{figure*}
    \centering
    \includegraphics[width=0.9\linewidth]{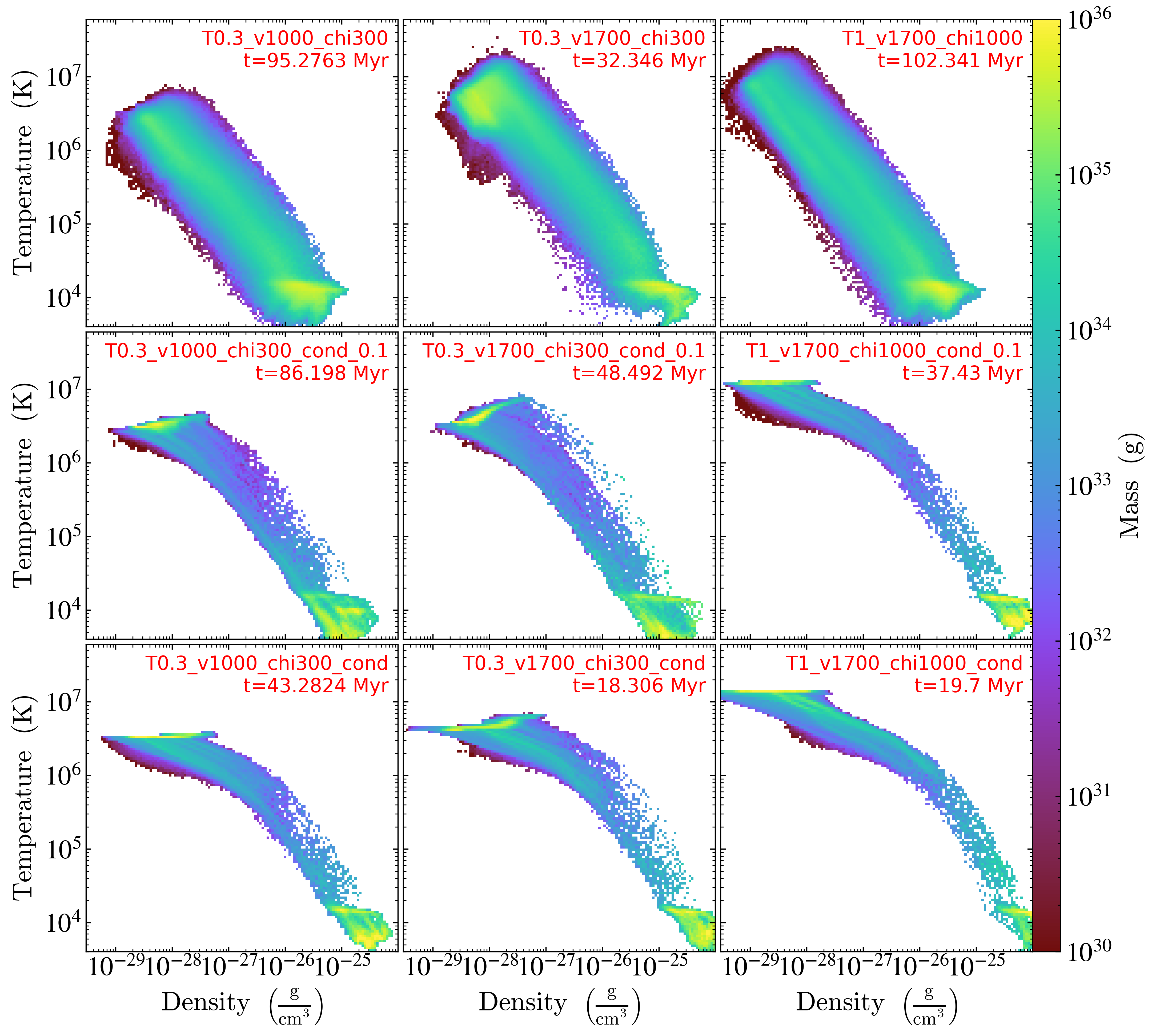}
    \caption{The phase space distributions of the cloud regions in several simulations (as labelled) at t50. \revision{The colour of each bin indicates the total mass of grid cells in the cloud region that have densities and temperatures within the bin. The physical times that these clouds have evolved are also marked in the figure.}}
    \label{fig:phase_space}
\end{figure*}

\yangedit{The effect of conduction can also be observed in the phase space distributions of the clouds. In Figure \ref{fig:phase_space}, we show three such examples. Here, the three columns correspond to simulations with three different sets of $(M,\chi_0)$ values, and in each column the conduction level increases from top to bottom. The cores of the clouds are clearly distinguishable in the lower right corner of all of the panels. The mixing layers in the non-conductive simulations all display very extended distributions in phase space, while the evaporated material in the conductive simulations is predominantly concentrated on the upper left.}

One might assume that the mass loss rate would be 10 times lower in the weak conduction runs as compared to the full conductive simulations, based on the suppression factor of $\eta=0.1$ (Equation \ref{eq:kappa}). However, this assumption is clearly not true for any of the simulations shown in Figure \ref{fig:mass_loss}. The reason for this discrepancy is that the suppression factor $\eta$ is only applied to the thermal conduction rate in the classical regime,  while the saturated conduction rate, which is solely determined by the total number density of free electrons, remains unaffected. Therefore, at the surface of the cloud, where the temperature gradient can be very steep, the thermal conduction rate is not reduced by a factor of 10 in the weak conduction runs compared to their full conduction counterparts. This is best demonstrated by the \revision{T3\_v3000\_chi3000\_cond\_0.1 and T3\_v3000\_chi3000\_cond} simulations, which have the steepest temperature gradient in all of our simulations. Apart from an early delay, we note that the clouds in \revision{both the weak and full conduction runs} have almost identical mass loss rates throughout their evolution, indicating that the thermal conduction across their surfaces is mostly saturated in both cases.

As discussed in \S\ref{subsec:simulations}, our clouds have lower column densities than those in the simulations conducted by \citet{Scannapieco2015} and \citet{Brueggen2016}. To demonstrate the effect of this, the mass loss curves of the T0.3\_v1700\_chi300 and T0.3\_v1700\_chi300\_cond simulations from their works, dubbed ``hcol'', are also shown in the top right panel of Figure \ref{fig:mass_loss}. The mass loss rate of our non-conductive T0.3\_v1700\_chi300 simulation is significantly higher than theirs. We find that this owes to inefficient cooling of the mixing layer in our simulations. Although the cooling time of the post-shock cloud center, which increases at lower densities, is still much shorter than the cloud crushing time in our T0.3\_v1700\_chi300 simulation (see Table \ref{tab:run_list}), we find that the cooling time of the warm gas in the mixing layer is much longer than the cloud crushing time. In the absence of conduction, this results in a much more extended mixing layer that dissolves into the ambient gas at a higher rate than in the high column density cloud. This only stops at $t\approx 8 t_{\mathrm{cc}}$, when the entire mixing layer is stripped away, significantly reducing the mass loss rate afterward. 

\revision{Likewise, the T0.3\_v1700\_chi300\_cond simulation also has a higher mass loss rate than its high column density counterpart. Unlike the non-conductive case, however, this results from a decrease in the evaporation time scale at lower column densities (see Equations 18 and 19 of \citealt{Brueggen2016}).}

Table \ref{tab:cloud_distance} shows $d_{25}$, the distance the clouds travel at t25 \revisiontwo{relative to the ambient flow}\revision{, as well as t25 itself}. We note that $d_{25}$ is over $100\ \mathrm{kpc}$ in several of our simulations. However, a few of our conductive clouds are notably short-lived and do not travel very far. For example, both the weakly conductive and fully conductive versions of T0.3\_v3000\_chi300 only travel $\approx 30\mathrm{kpc}$ at t25. As discussed above, this owes to fast conductive evaporation. On the other hand, in some simulations, such as T0.3\_v1700\_chi300, the clouds travel much farther than $d_{25}$ before fully dissolving, because their mass loss rates decrease significantly at low mass fractions. It should also be noted that the hot gas in the CGM does not have a uniform density distribution. As a cloud travels outward, the ambient density decreases over time. This may increase the lifetime and travel distance of the cloud.

\begin{table}
    \centering
    \caption{The distances the clouds travel relative to the ambient gas at t25.}
    \begin{tabular}{lcc}
        \hline
        Name & $d_{\mathrm{25}}/\mathrm{kpc}$ & \revision{$t_{25}/\mathrm{Myr}$}\\\hline
        T0.3\_v1000\_chi300 & \revisiontwo{125} & 139 \\
        T0.3\_v1000\_chi300\_cond\_0.1 & \revisiontwo{123} & 130 \\
        T0.3\_v1000\_chi300\_cond & \revisiontwo{63} & 71 \\
        T0.3\_v1700\_chi300 & \revisiontwo{68} & 44 \\
        T0.3\_v1700\_chi300\_cond\_0.1 & \revisiontwo{114} & 71 \\
        T0.3\_v1700\_chi300\_cond & \revisiontwo{65} & 42 \\
        T0.3\_v3000\_chi300 & \revisiontwo{142} & 49 \\
        T0.3\_v3000\_chi300\_cond\_0.1 & \revisiontwo{29} & 11 \\
        T0.3\_v3000\_chi300\_cond & \revisiontwo{17} & 7 \\
        T1\_v1700\_chi1000 & \revisiontwo{215} & 134 \\
        T1\_v1700\_chi1000\_cond\_0.1 & \revisiontwo{94} & 59 \\
        T1\_v1700\_chi1000\_cond & \revisiontwo{44} & 30 \\
        T3\_v3000\_chi3000 & \revisiontwo{234} & 81 \\
        T3\_v3000\_chi3000\_cond\_0.1 & \revisiontwo{66} & 25 \\
        T3\_v3000\_chi3000\_cond & \revisiontwo{49} & 19 \\
        T0.1\_v150\_chi100 & \revisiontwo{34} & 299 \\
        T0.1\_v150\_chi100\_cond\_0.1 & \revisiontwo{54} & 442 \\\hline
    \end{tabular}
    \label{tab:cloud_distance}
\end{table}

\section{Ion densities}
\label{sec:ion_densities}
\subsection{Ion maps}
\label{subsec:ion_maps}
 For this work, we include 14 ions in our analysis: H \textsc{i}, O \text{I}, Mg \textsc{ii}, Si \textsc{ii}, C \textsc{ii}, O \textsc{ii}, Si \textsc{iii}, C \textsc{iii}, Si \textsc{iv}, O \textsc{iii}, C \textsc{iv}, O \textsc{iv}, O \textsc{vi} and Ne \textsc{viii}, listed in order of increasing ionization potential. To obtain the ion densities of the clouds, we follow a similar approach to \citet{Cottle2018} that we will describe here. Assuming solar metallicities and the \citet{Haardt2012} (HM12) background at redshifts $z=0.101,0.540,1.053$, $2.013$ and $4.895$, we first generate ionization fraction tables using Cloudy 23 \citep{Chatzikos2023} \revision{under the assumption of ionization equilibrium}. Using these tables, we compute the ion densities of each grid cell at each of the redshifts listed above using the \textsc{yt}-based \citep{Turk2011} \textsc{trident} code \citep{Hummels2017} under the assumption that the cloud is optically thin and does not have significant self-shielding. The ion density distributions are then projected onto pixelated maps using \textsc{yt}'s off-axis projection function, with lines of sight at different angles $\theta$ from the direction of the ambient flow. For each simulation, we generate 11 such maps at projection angles of $\cos \theta$ ranging from 0 to 1 with a step of $0.1$ and use a pixel size of $10\ \mathrm{pc}$, which is $\approx 2$ times the finest resolution in our simulations.
 
\begin{figure*}
	\centering
 	\includegraphics[width=0.99\linewidth]{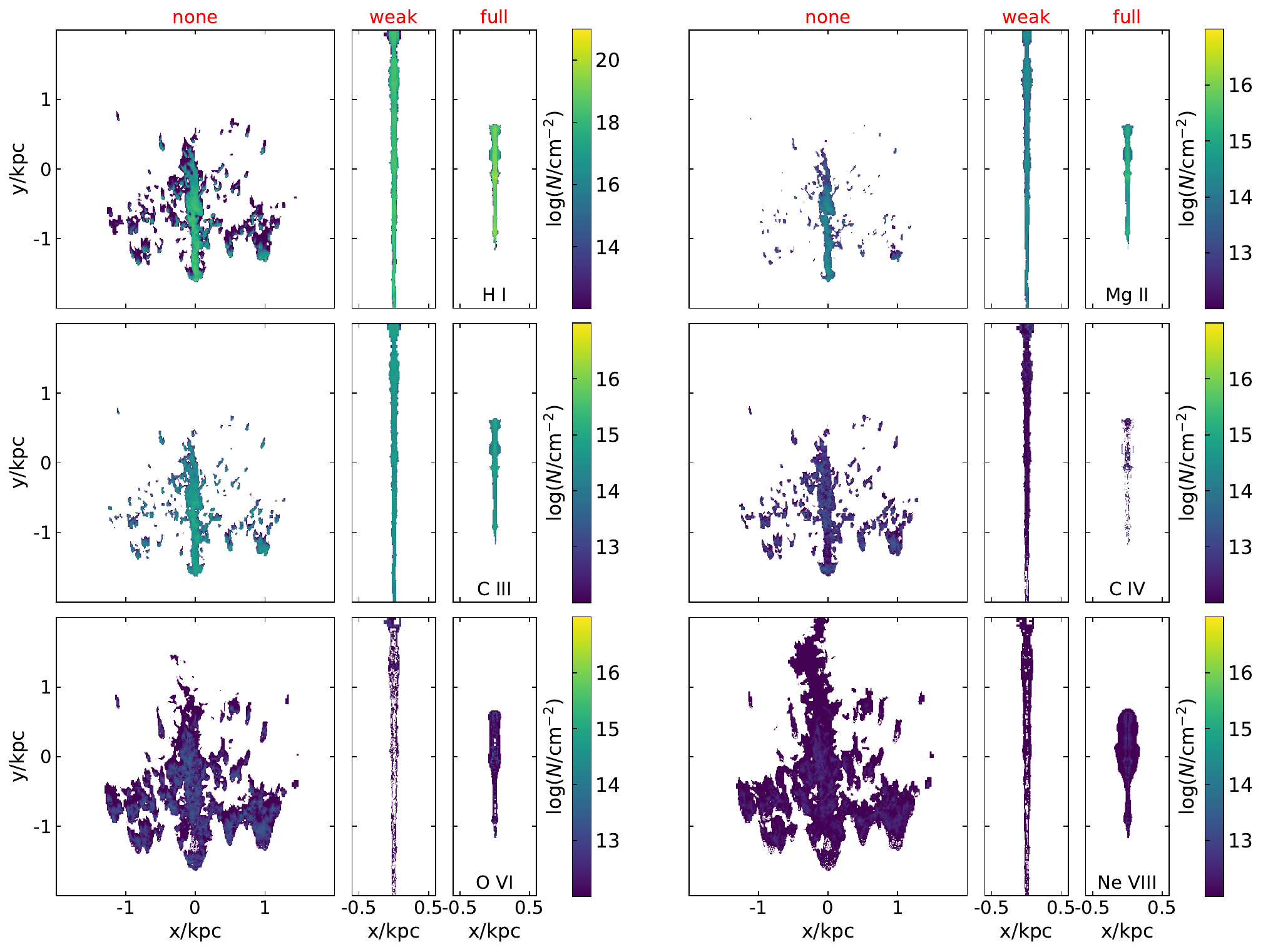}
 	\caption{Edge-on projections of the distributions of several ions in the cloud regions of the T0.3\_v1700\_chi300 cloud crushing simulations with 3 different conduction levels (none, weak and full) at t50 and $z=0.540$.}
 	\label{fig:ion_colden_maps_t2_z0.540}
\end{figure*}
 Figure \ref{fig:ion_colden_maps_t2_z0.540} shows the edge-on column density distributions of several ions in the cloud regions of the T0.3\_v1700\_chi300 simulations with different conduction levels at t50 with $z=0.540$ \revisiontwo{(hereinafter, when we refer to sets of simulations such as ``the T0.3\_v1700\_chi300 simulations", it is intended to include all 3 simulations with different conduction levels that share the referred ambient temperature and velocity)}.  From left to right, the three columns show simulations with no conduction, weak conduction, and full conduction. As discussed in \citet{Brueggen2016}, there is a clear morphological difference between the conductive and non-conductive clouds. The non-conductive cloud fragments, owing to Kelvin-Helmholtz instabilities (KHI), which is smoothed out by thermal conduction, leaving the conductive cloud largely intact. This occurs in all of the cloud crushing runs that we study in this work. In addition, we also note that in all three simulations, different ions show different distribution patterns. For example, low ions such as Mg \textsc{ii} are concentrated at the center of the clouds, while the higher ions spread out to the surface of the clouds as the ionization potential increases.
 
\begin{figure*}
	\centering
 	\includegraphics[width=0.7\linewidth]{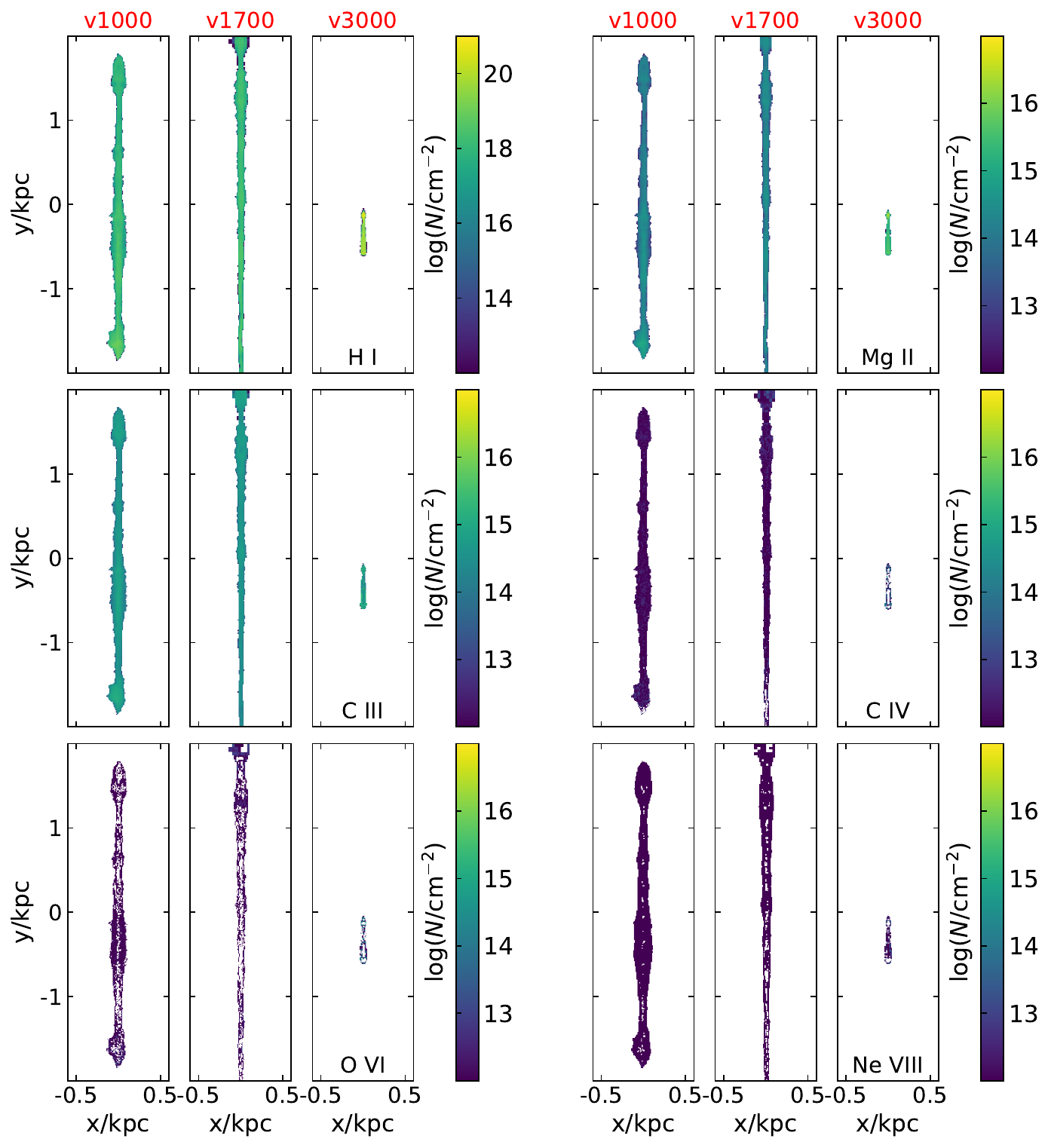}
 	\caption{Edge-on projections of the distributions of several ions in the T0.3\_v1000\_chi300\_cond\_0.1, T0.3\_v1700\_chi300\_cond\_0.1 and T0.3\_v3000\_chi300\_cond\_0.1 at t50 and $z=0.540$.}
 	\label{fig:ion_colden_maps_runs_t2_z0.540}
\end{figure*}
In Figure \ref{fig:ion_colden_maps_runs_t2_z0.540}, we compare the three $\chi=300$ weakly-conductive simulations with different Mach numbers at t50. From left to right, the Mach numbers are 3.8 ($v_{\mathrm{hot}}=1000\ \mathrm{km/s}$), 6.5 ($v_{\mathrm{hot}}=1700\ \mathrm{km/s}$) and 11.4 ($v_{\mathrm{hot}}=3000\ \mathrm{km/s}$). It is clear that the morphologies of the clouds with $M=3.8$ and $M=6.5$ are very similar, while the $M=11.4$ cloud is much more compressed. From Figure 7 of \citet{Brueggen2016}, their conductive cloud with $M=11.4$ and $\chi_0=300$, which has a higher column density than ours, does not display such a strong compression. This compression likely owes to inefficient cooling in the cloud, as the cooling time of the post-shock cloud is actually longer than the cloud crushing time in our simulation (see Table \ref{tab:run_list}). This causes rapid evaporation early on (see Figure \ref{fig:mass_loss}), which in turn compresses the cloud. A detailed analysis of the dependence of the cloud morphology on these various time scales is, however, beyond the scope of this work.
 
\subsection{Mock spectra and absorbers}\label{subsec:mock_spectra}
In the projected maps, each pixel corresponds to a line of sight. As a result of the complex and non-uniform nature of the clouds in our simulations, many of these pass through multiple distinct absorbers. Therefore, instead of using the total column density for our analysis, we fit individual lines. For each map, we randomly select 10\% of all pixels with a column density of any metal ion larger than $10^{11}\ \mathrm{cm^{-2}}$ to save computer time. We use the \texttt{LightRay} module of Trident to generate an optical depth spectrum of each ion for each pixel with a velocity resolution of $5\ \mathrm{km/s}$. If the peak optical depth, $\tau_{\mathrm{peak}},$ of an ion is larger than a threshold value of $10^{-4}$, we normalize the spectrum to a peak value of 4 by multiplying the entire optical depth spectrum by a factor of $k=4/\tau_{\mathrm{peak}}$, and generate a normalized mock absorption spectrum with $\bar{F}=\exp(-k \tau)$. 

\begin{figure*}
    \centering
    \includegraphics[width=0.9\linewidth]{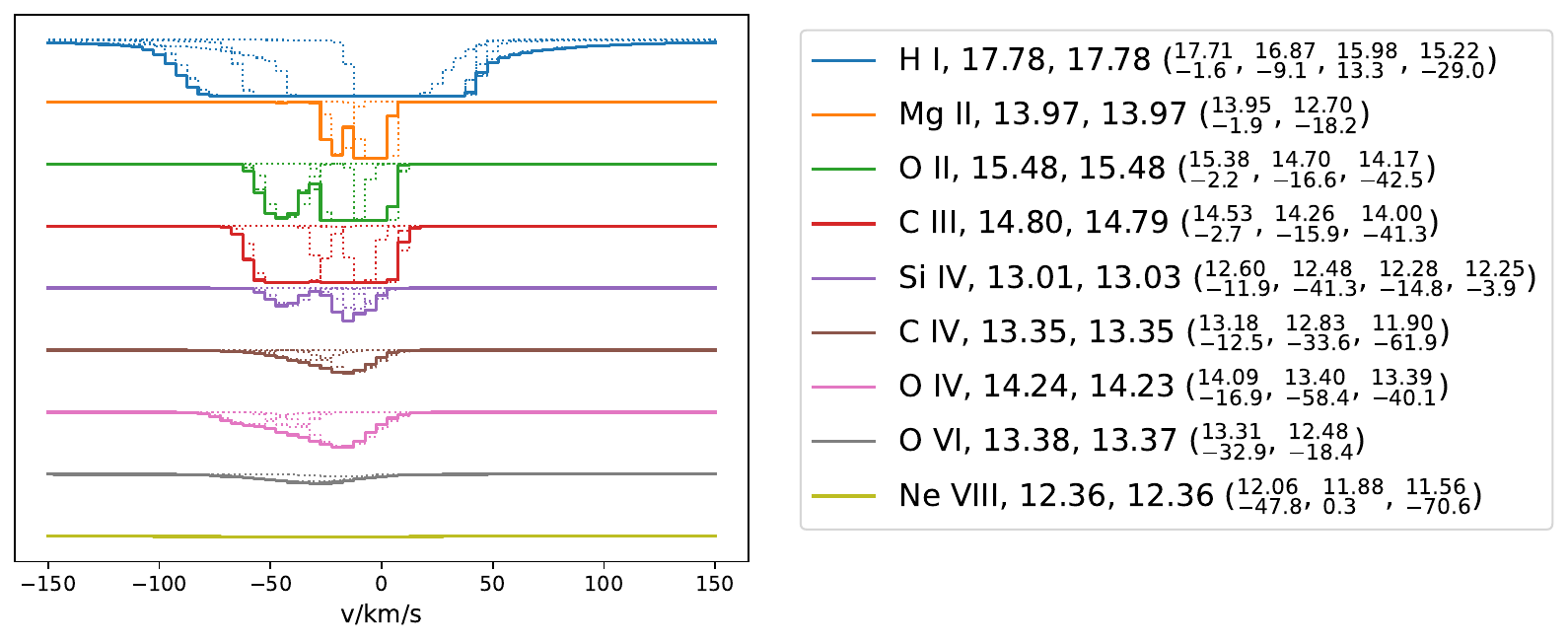}
    \caption{Absorption spectra of several ions from a line of sight through the T0.3\_v1700\_chi300 simulation. This specific line of sight is at an angle of $60^{\circ}$ from the direction of the ambient velocity and goes through the rear part of the cloud at a time when $\approx 50 \%$ of the cloud mass remains. The total fitted fluxes are marked by dashed lines, and the fitted fluxes of each identified component are marked by dotted lines. The first number after the ion name in the legend is the total $\log N$ for this specific line of sight, the second number is the total $\log N$ for all the fitted components, and the numbers in the parenthesis are \revision{the column density (top) and velocity offset (bottom)} of each individual fitted component.}
    \label{fig:absorption_example}
\end{figure*}

\begin{figure*}
    \centering
    \includegraphics[width=0.92\linewidth]{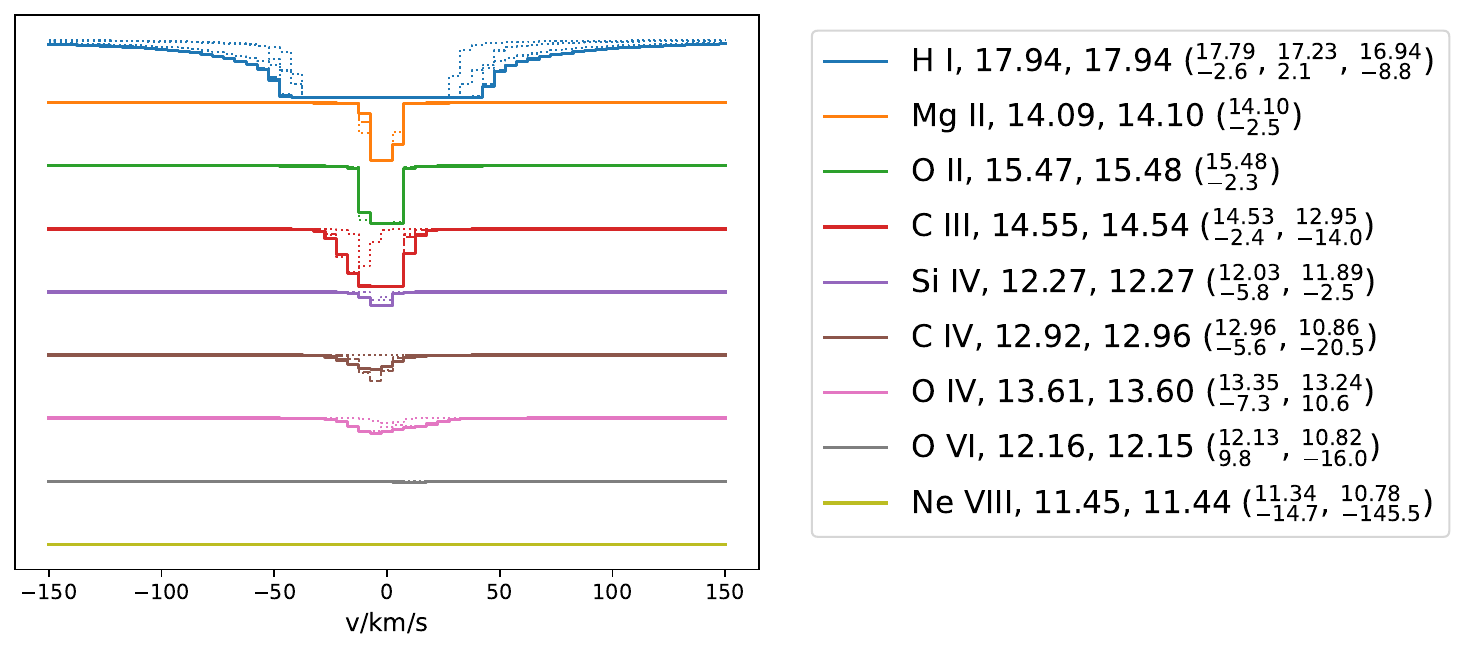}
    \caption{Absorption spectra from a line of sight through the T0.3\_v1700\_chi300\_cond\_0.1 simulation at an angle of $60^{\circ}$ from the direction of the ambient velocity at a time when $\approx 50 \%$ of the cloud mass remains. The annotations are the same as Figure \ref{fig:absorption_example}.}
    \label{fig:absorption_example_cond}
\end{figure*}

The normalized absorption spectrum is fed into the \texttt{absorption\_spectrum\_fit} module of Trident, which identifies individual \revision{absorption features}, fits them with Gaussian profiles, and gives a list of normalized column densities $\bar{N}$ and line widths $b$ of individual absorbers. \revision{The default tolerance for the Trident spectrum fitter is $10^{-4}$, which we find performs well for metal ions in accounting for all absorption features while avoiding overfitting (see Figures \ref{fig:absorption_example} and \ref{fig:absorption_example_cond}). For H \textsc{i}, however, the optical depths are often several orders of magnitude larger than 1, resulting in very small normalization factors that can cause secondary absorption features that are clearly visible in the unnormalized spectra to be missed when fitting the normalized ones. To avoid this, when fitting the H \textsc{i} spectra, we lower the tolerance to $10^{-7}$.} We then denormalize these column densities using $N=\bar{N}/k$ to obtain the real column densities of the absorbers, while the line widths remain unchanged. 

Because of the normalization of the optical depths, our lists of absorbers do not always reflect what can actually be fit in real observations. Our goal here is to separate absorbers that are physically distinct, regardless of whether or not they are distinguishable observationally.

Apart from the aforementioned physical motivation, another advantage of our approach is its flexibility with regard to scaling. In observations, the clouds encountered may have different masses, sizes, and metallicities than those in our simulations. In addition, the UV background responsible for the photoionization can be different from the ones we use. At higher column densities, self-shielding may also become important. As a zeroth-order approximation, we can scale our clouds to match these observations by multiplying the column densities of our absorbers by constant factors, allowing absorbers that are below the observational limit to become detectable. It is for the same reason that when we study the column density distributions of our absorbers in \S \ref{sec:column_densities}, we include all of the absorbers with column densities down to $10^{11}\ \mathrm{cm^{-2}}$, which is far below the detection limit for any of our ions. After scaling, these column density histograms can be shifted horizontally so that absorbers at the low column density end become observable.

While we include photoionization in our ion fraction calculations, photoheating is not included in the simulations. Through photoheating, the temperature of the low-density, cool regions with $T\approx 10^4\ \mathrm{K}$ can be raised to several times $10^4\ \mathrm{K}$, which could potentially affect the distributions of the low ions. To study the effect of this, we artificially set a temperature floor of $T_{\mathrm{f}}=3\times10^4\ \mathrm{K}$ on all grid cells with densities $\rho<\rho_{\mathrm{c}}/3=3.3\times 10^{-27}\ \mathrm{g/cm^3}$. We find that this does not affect the distribution of any ion in our simulations. Photoheating is therefore not important in our analysis.

Figure \ref{fig:absorption_example} showcases one set of mock spectra from the T0.3\_v1700\_chi300 simulation. The zero point of the velocity is set to the H \textsc{i}-mass-weighed line-of-sight velocity. The unnormalized absorption spectra are shown as solid lines. For each ion, we also show the absorption spectrum of each fitted component as a dotted line, as well as the total fitted spectrum as a dashed line. In the legend, the name of each ion is followed by a set of numbers. The first number is the real $\log N$ of the ion, the second number is $\log(\Sigma N)$ for all the fitted components, and the numbers in the parenthesis are the $\log N$ for each individual fitted component. It is clear that this system consists of several distinct absorbers. Interestingly, the component on the left, centered at $\approx -50\ \mathrm{km/s}$, only appears in the middle ions. We find that this is a common occurrence in the spectra of the non-conductive simulations, where such components originate from fragments that are separated from the central core.

Because of the relative integrity of the clouds, the conductive runs usually either do not display multiple components or have components that are very close to each other in their spectra. Figure \ref{fig:absorption_example_cond} shows an example from the T0.3\_v1700\_chi300\_cond\_0.1 simulation, where there is only one distinguishable component across all ions up to O \textsc{iv}.

\section{Column densities and line widths of the absorbers}\label{sec:column_densities}
\subsection{Mach number, density contrast and conduction}
\begin{figure*}
    \centering
    \includegraphics[width=0.9\linewidth]{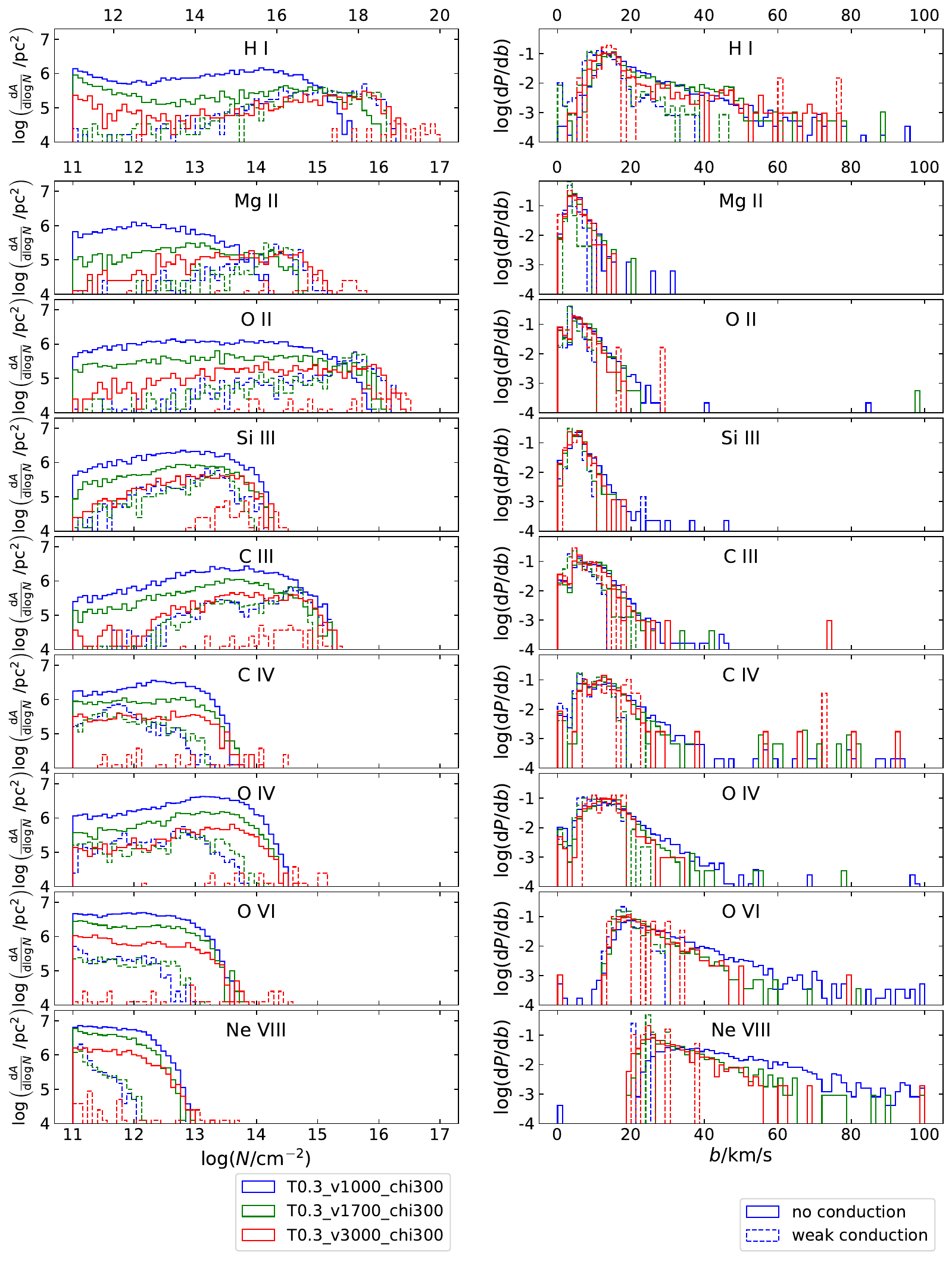}
    \caption{A comparison of the column density (left column) and line width (right column) distributions of the absorbers in cloud crushing simulations with $\chi=300$, $z=0.540$ at t50. The lines of sight are at an angle of $60^{\circ}$ from the direction of the ambient flow. The line styles indicate the level of conduction. The area under each column density histogram is equal to the covering area of the corresponding ion, while that under each line width histogram equals 1 \revision{(hence labeled $\mathrm{d}P/\mathrm{d}b$ as probability densities)}. The line width distributions only include absorbers with $N_{\mathrm{ion}}>10^{12}\mathrm{cm^{-2}}$ for the corresponding ions. From top to bottom, the ions are ranked in order of increasing ionization potential.}
    \label{fig:compare_mach_t2_5_z0.540}
\end{figure*}
In Figure \ref{fig:compare_mach_t2_5_z0.540}, we show the column density and line width distributions of the absorbers in several cloud crushing simulations with $\chi=300$, $z=0.540$ at t50, and with lines of sight at an angle of $60^{\circ}$ from the direction of the ambient flow. The solid lines indicate the non-conductive simulations, and the dashed lines show the weakly-conductive simulations. In the non-conductive simulations, increasing the Mach number results in fewer low-column density absorbers across all ions. At the high column density end, the number of absorbers increases with the Mach number for the low ions H \textsc{i}, Mg \textsc{ii} and O \textsc{ii}. For ions with higher ionization potentials, however, the number of high column density absorbers is not strongly affected by the Mach number.

In comparison to their non-conductive counterparts, the weakly-conductive simulations have more high column density absorbers of low ions, such as H \textsc{i} and Mg \textsc{ii}. However, the effect of conduction on the high ions depends on the Mach number. While the run with the highest Mach number, T0.3\_v3000\_chi300\_cond\_0.1, produces more high ion absorbers on the high column density end than T0.3\_v3000\_chi300, the other two weakly-conductive runs with smaller Mach numbers show fewer high ion absorbers than their non-conductive counterparts across the entire column density range. This owes to the extreme compression of the T0.3\_v3000\_chi300\_cond\_0.1 cloud (see Figure \ref{fig:ion_colden_maps_runs_t2_z0.540}).

\yangedit{The right panel of Figure \ref{fig:compare_mach_t2_5_z0.540} shows the line width distributions of these simulations. Here we only include absorbers with column densities $N>10^{12}\ \mathrm{cm^{-2}}$ to avoid having the line width distributions dominated by absorbers with lower column densities, which are not observable unless we have a very large scaling factor.} We find that the non-conduction runs have more broad absorbers than the weakly-conductive runs for all the ions. This likely owes to the fragmentation of the clouds in the non-conductive runs, which results in more kinematic broadening, whereas the conductive clouds remain intact due to the suppression of KHI by thermal conduction. 

\begin{figure*}
	\centering
	\includegraphics[width=0.9\linewidth]{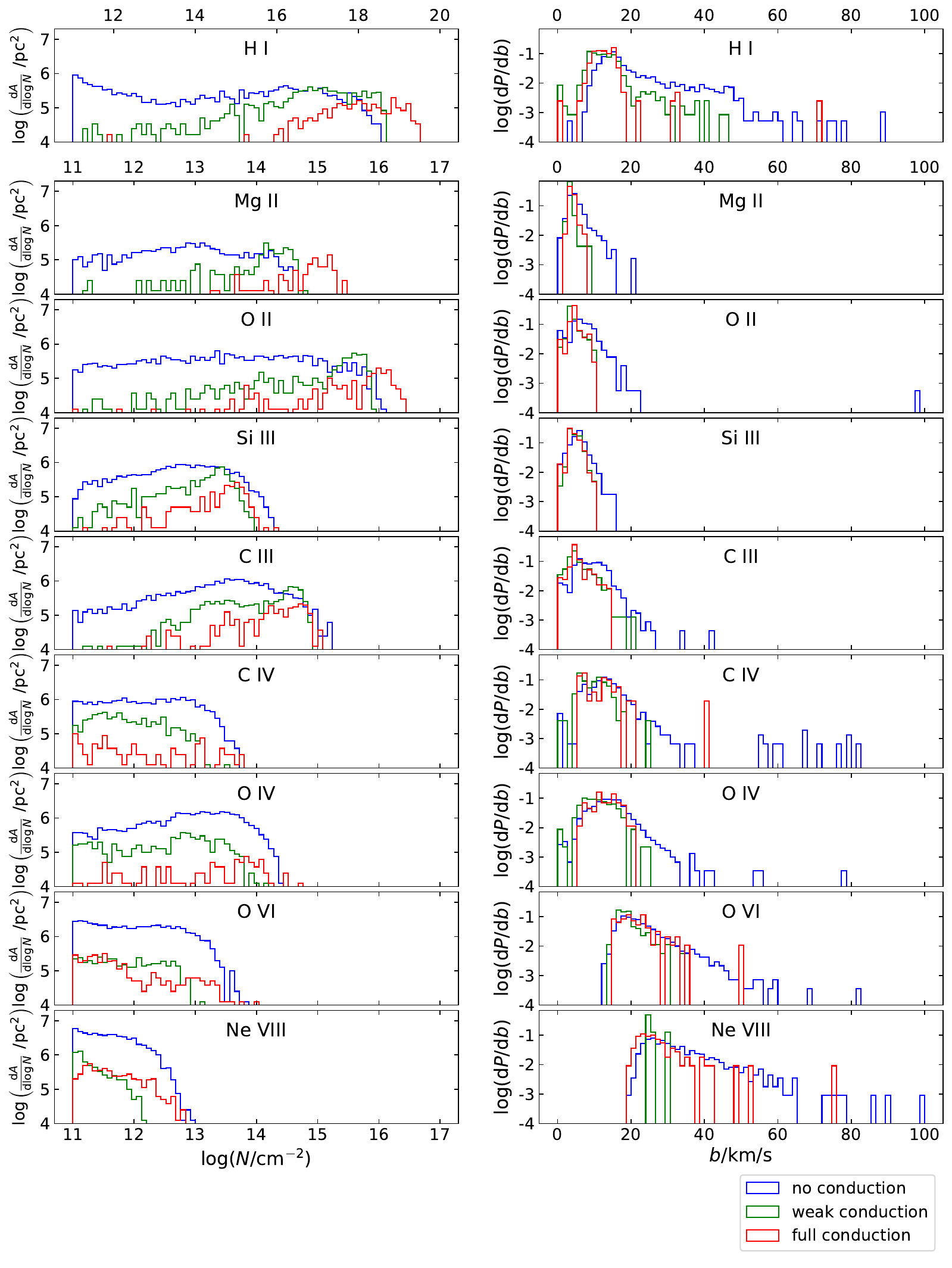}
	\caption{The column density (left column) and line width (right column) distributions of the absorbers in the T0.3\_v1700\_chi300 simulations with 3 different levels of conduction: no conduction (blue), weak conduction (green) and full conduction (red). The redshift of the radiation background, the evolutionary time, and the line-of-sight angle are the same as those used for Figure \ref{fig:compare_mach_t2_5_z0.540}.}
	\label{fig:compare_conduction_T0.3_v1700_chi300_ld2_t2_5_z0.540}
\end{figure*}

To further explore how thermal conduction affects the ions, we compare the \revision{column density and line width distributions of ions in the} T0.3\_v1700\_chi300 simulations with 3 different conduction levels (no conduction, weak conduction, and full conduction) in Figure \ref{fig:compare_conduction_T0.3_v1700_chi300_ld2_t2_5_z0.540}. From this figure, it is clear that the level of conduction has a profound effect on the distributions of the ions. Compared to the weakly conductive simulation, we note that the cloud in the fully conductive simulation has more high column density absorbers for almost all ions. This is an effect of the stronger evaporative compression in the fully conductive run, as can be observed in Figure \ref{fig:ion_colden_maps_t2_z0.540}. \revision{Comparing the line width distributions, we note that ions with ionization potentials between those of C \textsc{iv} and O \textsc{vi} have significantly higher cutoffs at the lower end than their non-conductive and weakly conductive counterparts. This is likely caused by the more efficient heating of the evaporative layer through thermal conduction.}

\begin{figure*}
    \centering
    \includegraphics[width=0.9\linewidth]{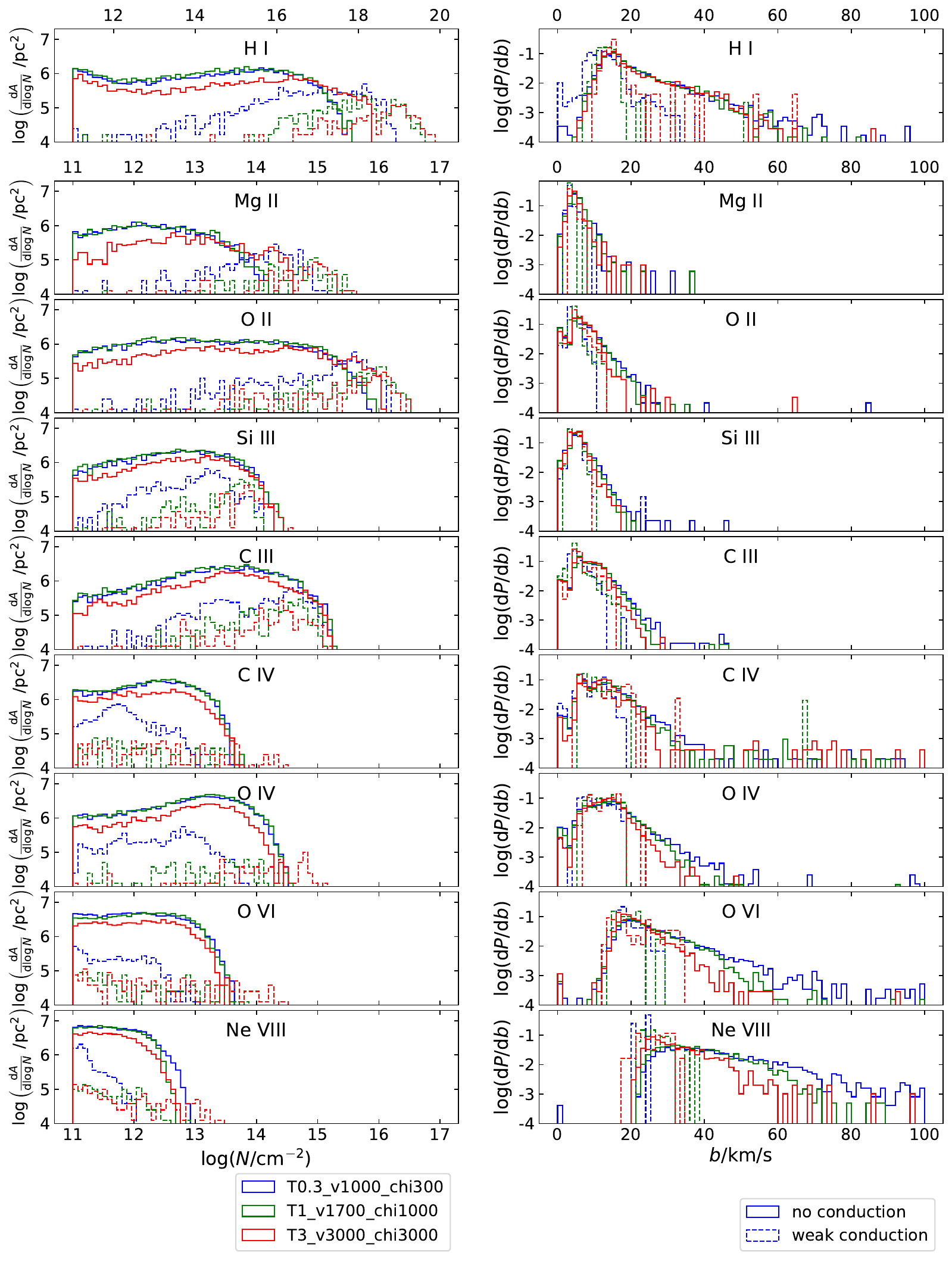}
    \caption{A comparison of the column density (left column) and line width (right column) distributions of the absorbers in several cloud crushing simulations with $M\approx3.5$, $z=0.540$ but with different density contrasts at t50. The lines of sight are at an angle of $60^{\circ}$ from the direction of the ambient flow.}
    \label{fig:compare_chi_t2_5_z0.540}
\end{figure*}
In Figure \ref{fig:compare_chi_t2_5_z0.540}, we show the ion density and linewidth distributions of simulations at similar Mach numbers $M\approx 3.5$ but with different density contrasts. In the non-conductive simulations, we find that the ion distributions do not depend strongly on the density contrast apart from an increase in the number of high column density absorbers of H \textsc{i}, Mg \textsc{ii} and O \textsc{ii} at the highest density contrast of 3000. This is consistent with the finding of \citet{Scannapieco2015}, that the scatter between different clouds with the same Mach number but different $\chi_0$ values is small. In the presence of thermal conduction, the impact of the density contrast on the ion distributions is more complex and ion-dependent. Nevertheless, there is an overall shift in column densities from low to high with increasing density contrasts for most ions, possibly because of stronger evaporative compression.

\subsection{Low $M_{\mathrm{hot}}$ and $\chi_0$}
\begin{figure*}
	\centering
	\includegraphics[width=0.9\linewidth]{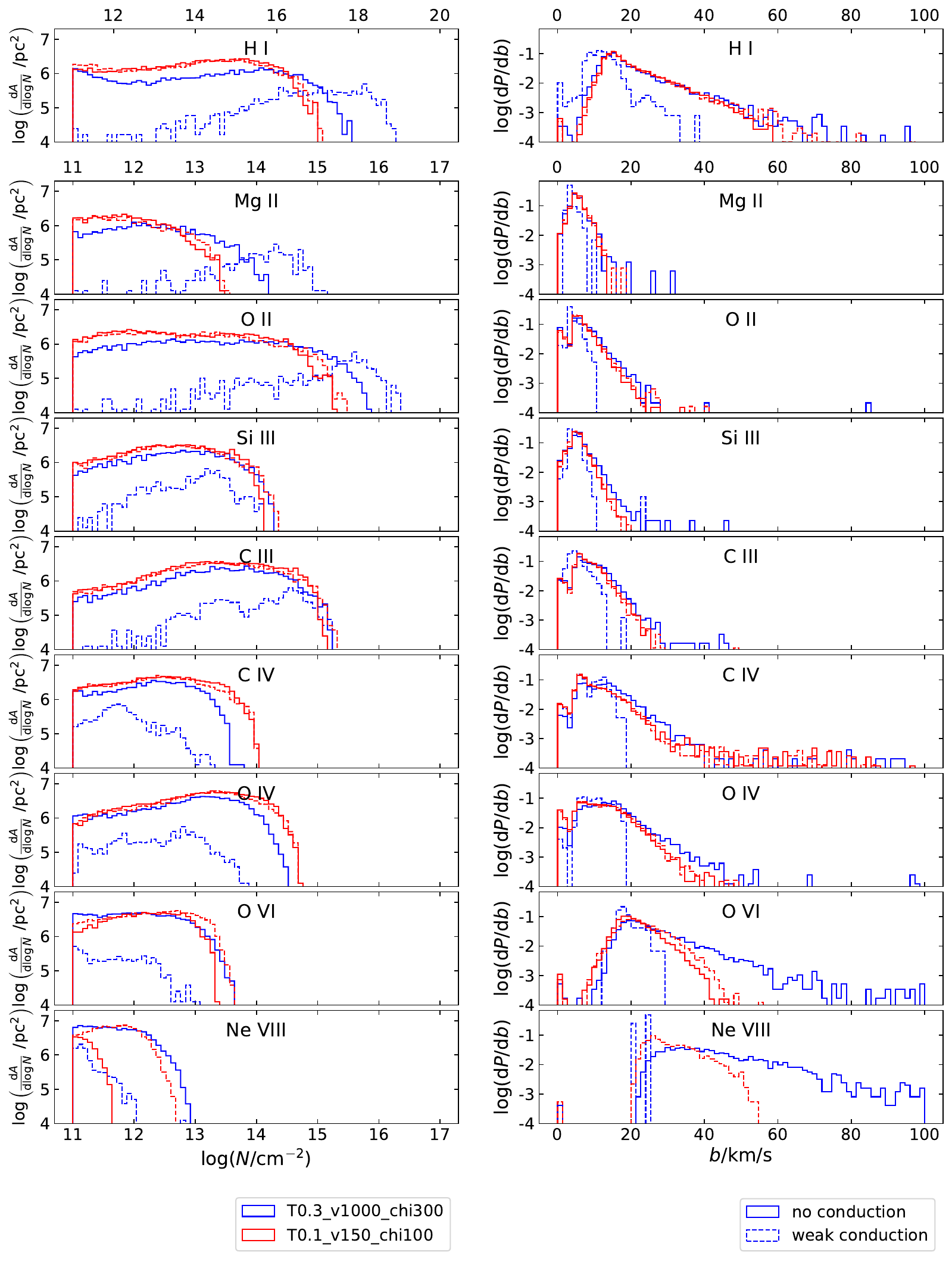}
	\caption{The column density (left column) and line width (right column) distributions of the absorbers in T0.1\_v150\_chi100 and T0.1\_v150\_chi100\_cond\_0.1. As a comparison, the clouds with the second lowest Mach number and density contrast, T0.3\_v1000\_chi300 and T0.3\_v1000\_chi300\_cond\_0.1 are also shown. The redshift and projection angle are the same as those in Figure \ref{fig:compare_mach_t2_5_z0.540}. The times are at t50 for all simulations.}
	\label{fig:phew_runs_T0.1_v150_chi100_cond_0.1_ld2_t2_5_z0.540}
\end{figure*}
To explore clouds with very low Mach numbers and density contrasts, \revision{in Figure \ref{fig:phew_runs_T0.1_v150_chi100_cond_0.1_ld2_t2_5_z0.540}} we compare the ion distributions of T0.1\_v150\_chi100 and T0.1\_v150\_chi100\_cond\_0.1 with those of T0.3\_v1000\_chi300 and T0.3\_v1000\_chi300\_cond\_0.1, which have the second lowest Mach numbers and density contrasts in our simulations. Because of the lower temperature gradient resulting from the lower density contrast, the thermal conduction in the T0.1\_v150\_chi100\_cond\_0.1 simulation is much weaker than in any of our other conductive simulations. This is reflected in the ion distributions: the column density and line width distributions are almost identical between T0.1\_v150\_chi100 and T0.1\_v150\_chi100\_cond\_0.1 for the low and middle ions, while only the distributions of O \textsc{vi} and Ne \textsc{viii} are significantly affected by the thermal conduction. 

Compared to the T0.3\_v1000\_chi300 clouds, the T0.1\_v150\_chi100 clouds have fewer high column density absorbers of H \textsc{i}, Mg \textsc{ii} and O \textsc{ii}, while producing more high column density absorbers of C \textsc{iv} and O \textsc{iv}. These are caused by a more extended mixing layer. Interestingly, in the cases of O \textsc{vi} and Ne \textsc{viii}, where thermal conduction does make a significant difference in the T0.1\_v150\_chi100 simulations, its effect on the ion distributions is the opposite from the T0.3\_v1000\_chi300 simulations. Namely, thermal conduction increases the numbers of absorbers of these ions in the former simulations, while it decreases them in the latter. This likely owes to a difference in the temperature of the evaporated material.

\subsection{Time evolution}
\begin{figure*}
    \centering
    \includegraphics[width=0.9\linewidth]{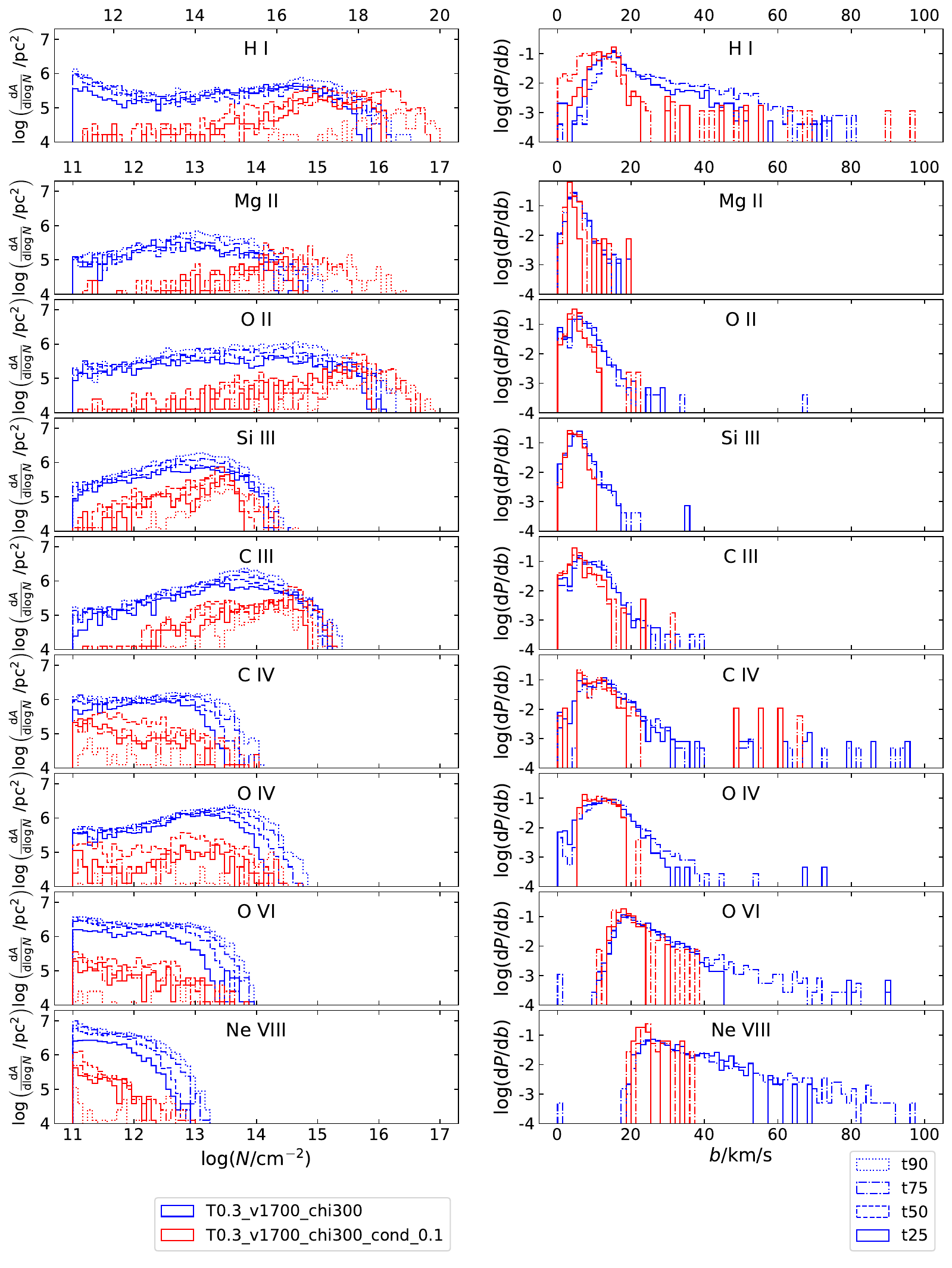}
    \caption{Time evolution of the column density (left column) and line width (right column) distributions of the absorbers in T0.3\_v1700\_chi300 and T0.3\_v1700\_chi300\_cond. The redshift and projection angle are the same as those in Figure \ref{fig:compare_mach_t2_5_z0.540}. The times, parameterized by the remaining mass fractions, are marked with different line styles. \revision{To avoid the lines overlapping excessively, only t75 and t25 are plotted for the line width distributions.}}
    \label{fig:absorber_distributions_5_z0.540}
\end{figure*}
Figure \ref{fig:absorber_distributions_5_z0.540} shows the time evolution of the column density and line width distributions of the absorbers in the T0.3\_v1700\_chi300 and T0.3\_v1700\_chi300\_cond\_0.1 simulations. The line style indicates the remaining mass fraction of the cloud, which we use as a proxy for the cloud's evolutionary stage. We first note that the non-conductive simulation shows very little evolution in the shape of the column density and line width distributions of all ions, with only an overall decrease in the numbers of absorbers across the entire column density range for each ion, as expected when the cloud loses mass. On the other hand, the evolution of the conductive simulation is more interesting. The evaporative compression, which finishes at around t90 in the T0.3\_v1700\_chi300\_cond\_0.1 simulation, drastically increases the number of high column density absorbers momentarily for the low ions, while the middle and high ions are much less affected, due to the efficient cooling within the dense compressed cloud.

\subsection{Projection angles}
\begin{figure*}
	\centering
	\includegraphics[width=0.9\linewidth]{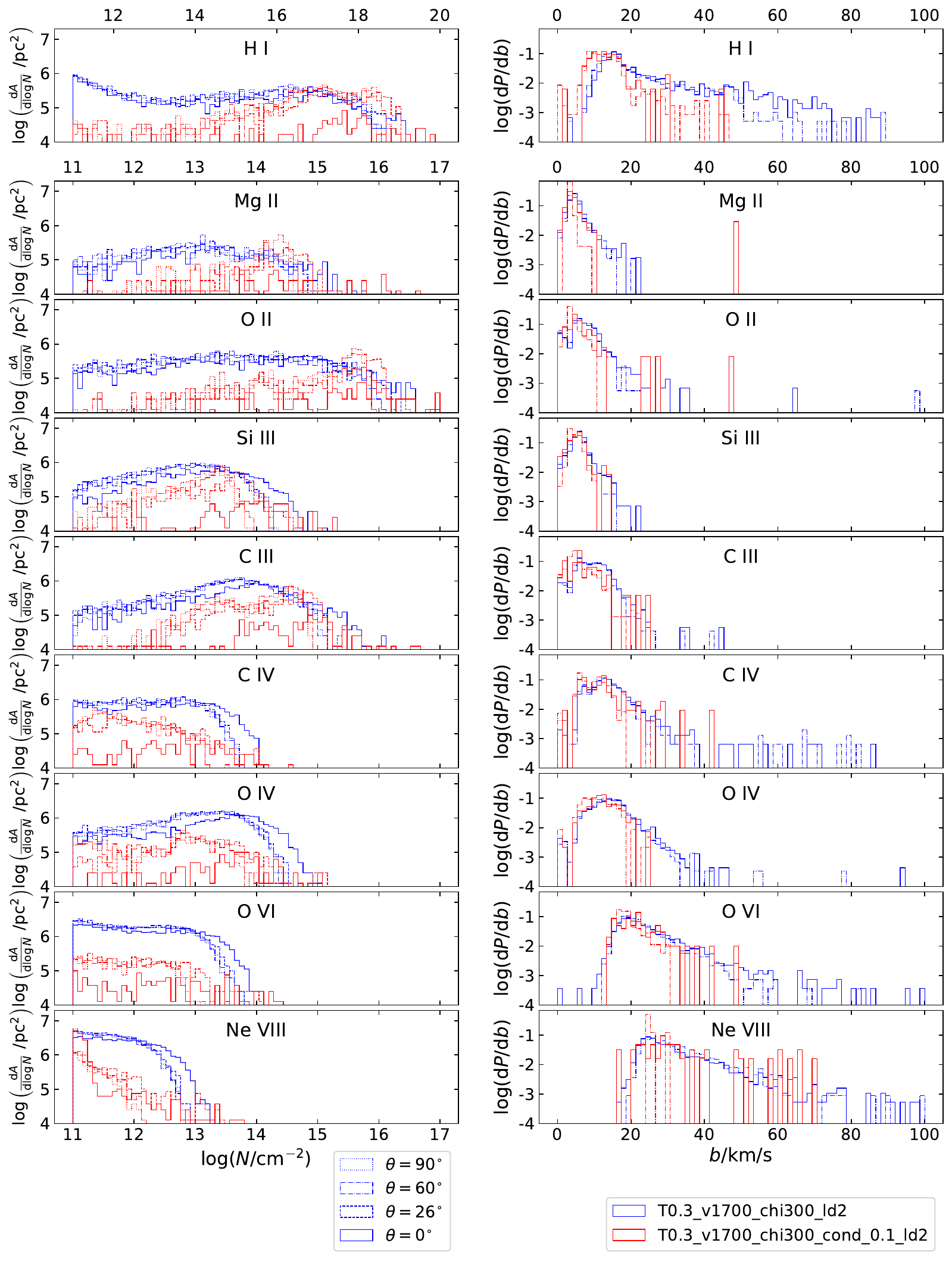}
	\caption{The column density and line width distributions of absorbers in the T0.3\_v1700\_chi300 and T0.3\_v1700\_chi300\_cond\_0.1 simulations at z=0.540 with 4 different projection angles. \revision{Here $\theta=0^{\circ}$ corresponds to the head-on sightline, while $\theta=90^{\circ}$ means edge-on. To avoid overlapping lines, only $\theta=60^{\circ}$ and $\theta=0^{\circ}$ are included for the line width distributions.} }
	\label{fig:compare_direction_t2_z0.540}
\end{figure*}
Figure \ref{fig:compare_direction_t2_z0.540} shows the column density and line width distributions of absorbers in the T0.3\_v1700\_chi300 and T0.3\_v1700\_chi300\_cond\_0.1 simulations at 4 different projection angles.  From this figure, It is immediately obvious that between $\theta=90^{\circ}$ (edge-on projections) and $\theta=26^{\circ}$ ($\cos^{-1}0.9$), the column density and line width distributions do not depend on the projection angle in the non-conductive simulation. It is only at a head-on angle of $\theta=0^{\circ}$ that we see much more high column density absorbers, as expected from the needle-like shape of the cloud. In the conductive simulation, there is a gradual increase in the column densities of absorbers with decreasing $\theta$. However, for most ions there is still a clear distinction between the head-on projection at $\theta=0^{\circ}$ and the other three angles. We find that all of our simulations display similar behavior to the ones shown in Figure \ref{fig:compare_direction_t2_z0.540}. This might indicate a significant difference in observed cloud properties between down-the-barrel observations and QSO absorption systems, where the clouds are unlikely to be perfectly aligned with the lines of sight.

\subsection{Redshift and photoionization}

\begin{figure*}
    \centering
    \includegraphics[width=0.9\linewidth]{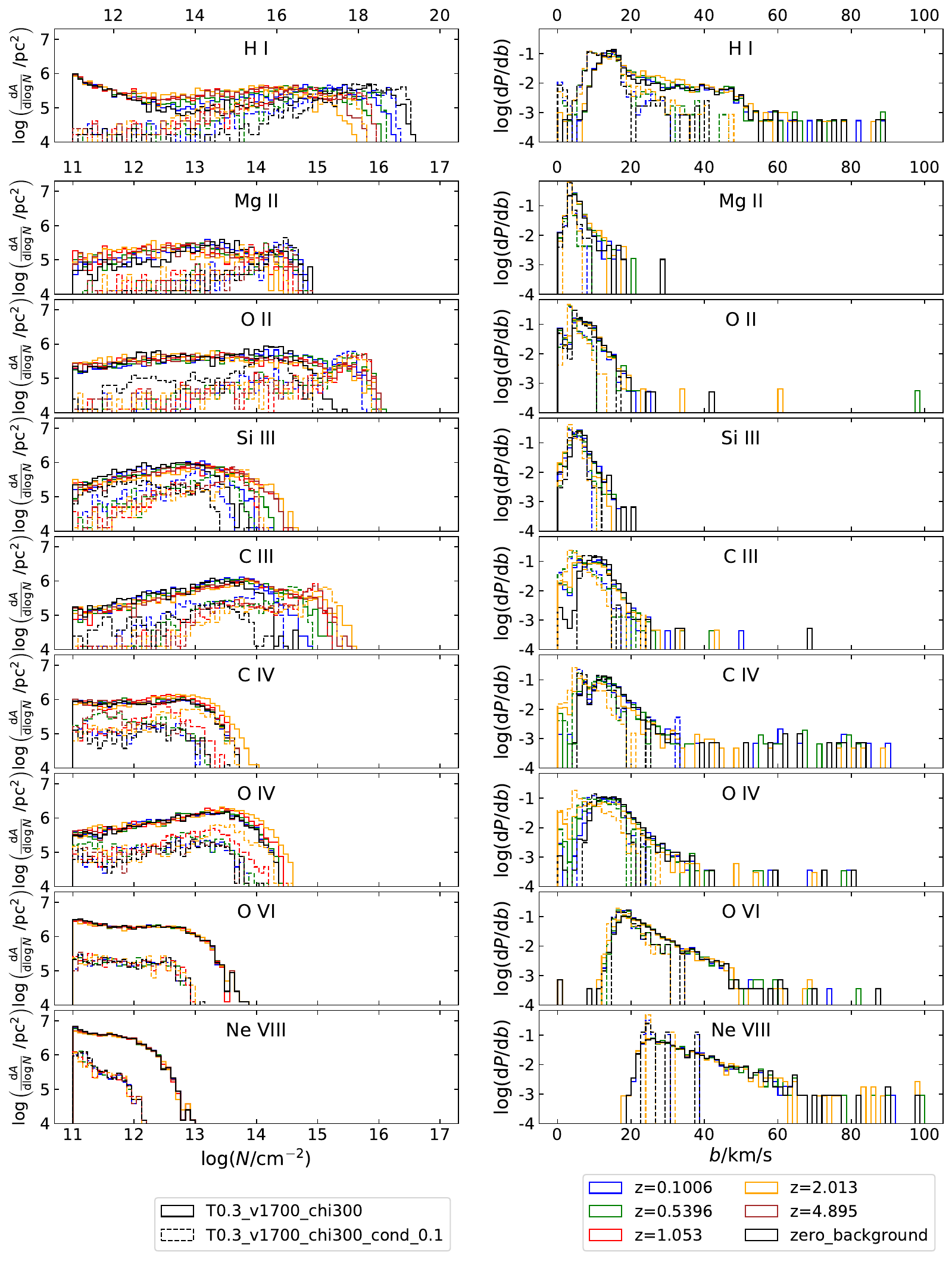}
    \caption{The column density (left column) and line width (right column) distributions of the absorbers in T0.3\_v1700\_chi300 and T0.3\_v1700\_chi300\_cond\_0.1, photoionized by the HM12 UV background at different redshifts, as well as without photoionization (``zero background"). The projection angles are $\theta = 60^{\circ}$. \revision{To avoid overlapping lines, $z=1.053$ and $z=4.895$ are not plotted for the line width distributions.}}
    \label{fig:compare_redshift_T0.3_v1700_chi300_cond_0.1_ld2_t2_5}
\end{figure*}

The role photoionization plays in QSO absorption systems has long been recognized and studied in observations (e.g. \citealt{Tripp2008, Chisholm2016}), simulations (e.g. \citealt{Oppenheimer2018, Appleby2022}) and theoretical calculations (e.g. \citealt{Strawn2022}). Because of the dependence of the UV radiation background on the redshift, the photoionization rates of the ions are also redshift-dependent. In Figure \ref{fig:compare_redshift_T0.3_v1700_chi300_cond_0.1_ld2_t2_5}, we show the absorber column density and line width distributions of the clouds in T0.3\_v1700\_chi300 and T0.3\_v1700\_chi300\_cond\_0.1 that are photoionized by the HM12 UV background at several different redshifts between $z=0.1$ and $z=5$. As a comparison, we also show the distributions of the same clouds under a zero background, where the ions are only collisionally ionized. 

The variation of the HM12 UV background with redshift is shown on the top panel of Figure 4 in \citet{Strawn2022}. Going up the redshift ladder, the background increases across the entire energy range between $z=0$ and $z=2$. From $z=2$ to $z=4$, the spectrum decreases slightly between the H \textsc{i} Lyman limit and the He \textsc{\textsc{ii}} Lyman series, but dips much further at energies higher than the He \textsc{\textsc{ii}} Lyman limit, which lies between the ionization energies of C \textsc{iii} and C \textsc{iv}. 

Figure \ref{fig:compare_redshift_T0.3_v1700_chi300_cond_0.1_ld2_t2_5}, shows the impact of photoionization on the ion distributions. The low ions H \textsc{i} and Mg \textsc{ii} are destroyed by photoionization, with the zero background having the largest number of high column density absorbers. For the higher ions O \textsc{ii}, Si \textsc{iii} and C \textsc{iii}, the numbers of high column density absorbers of these ions are much larger with photoionization in both the conductive and non-conductive simulations. This indicates that these ions are predominantly photoionized. The actual dependence on redshift varies between ions. For example, in both simulations the numbers of O \textsc{ii} absorbers on the high column density end are only weakly dependent on redshift, possibly owing to charge exchange with H \textsc{ii}. Meanwhile, for Si \textsc{iii} and C \textsc{iii} the numbers of high column density absorbers increase significantly with redshift between $z=0.101$ and $z=2.013$, and decrease slightly at $z=4.895$. This corresponds with the increase and decrease of the flux of UV photons from the HM12 background at these redshifts. Photoionization contributes much less for the higher ions. For C \textsc{iv} and O \textsc{iv}, photoionization is only significant between $z=1.053$ and $z=2.013$. Furthermore, O \textsc{vi} and Ne \textsc{viii} are almost entirely collisionally ionized.

Comparing the line width distributions, we note that an increase in photoionization correlates with an increase in the number of narrow absorbers \revision{for the middle ions ranging from Si \textsc{iii} to O \textsc{iv}}. We attribute this to the photoionized absorbers having lower temperatures than the collisionally ionized ones.

\subsection{Correlations between different ions}
In observed QSO absorption systems, ions with different ionization potentials are frequently found to be well aligned. From Figures \ref{fig:absorption_example} and \ref{fig:absorption_example_cond}, it is clear that such alignment occurs in our simulations as well. To investigate this systematically, we devise a way to identify matching absorbers of different ions from each line of sight. In this part of our study, we only include absorbers of each ion with $N_{\mathrm{ion}}>10^{12}\ \mathrm{cm^{-2}}$. For each pair of ions, which we refer to as ion 1 and ion 2, we go through the fitted absorber list of ion 1 in order of increasing velocity offsets. For each ion 1 absorber, we go through the absorber list of ion 2 in the same order until we find an absorber within $\Delta v=10\ \mathrm{km/s}$ of the ion 1 absorber. The pair is now considered matched, and the ion 2 absorber is removed from our search list. This process is repeated for all ion 1 absorbers. 

While it is easy to prove that the pairs matched this way remain unchanged if ion 1 and ion 2 are switched, if one of the ions has two absorbers that are both within $10 \mathrm{km/s}$ of an absorber of the other ion, the matching process is always biased toward the one with a lower velocity offset. To mitigate this, we repeat the entire search process in the reverse order in velocity space (i.e. going from high to low velocity offsets). The average number of matched pairs that are found in both orders is subsequently taken as the actual number of matched pairs.

\begin{figure*}
	\centering
	\includegraphics[width=0.99\linewidth]{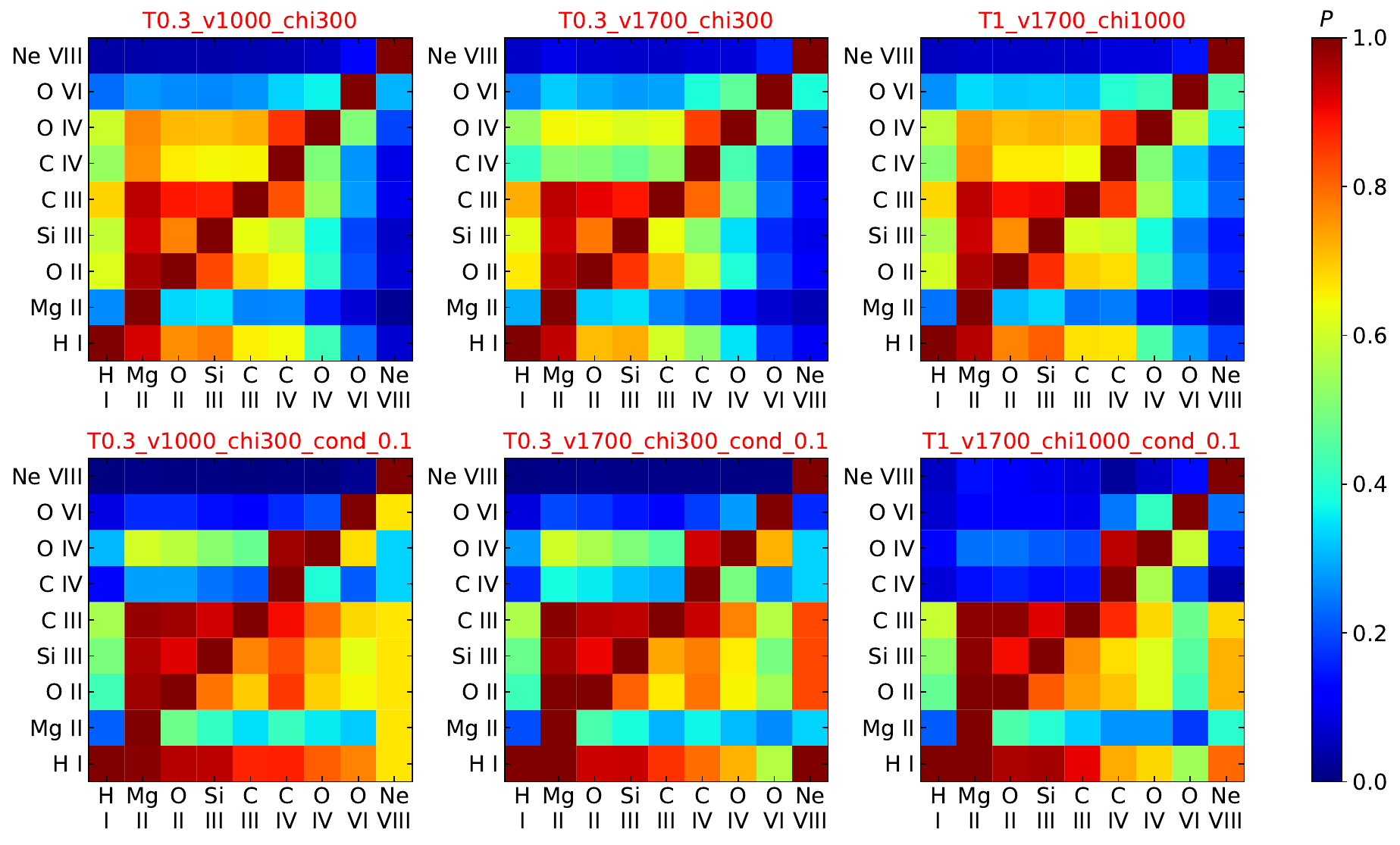}
	\caption{The matching probabilities of ions in 3 of our non-conductive cloud-crushing simulations (top row) and their weakly conductive counterparts (bottom row) at t50, $z=0.540$ and a projection angle of $\theta=60^{\circ}$ from the direction of the ambient flow. From left to right, the Mach numbers and initial density contrasts of the runs in each column are $(M,\chi_0)=(3.8,300)$, $(6.5,300)$ and $(3.5,1000)$, respectively. The colour of each grid cell indicates the fraction of absorbers of the ion on the $x$-axis that is matched to those of the ion on the $y$-axis.}
	\label{fig:matching_probability}
\end{figure*}

\yangedit{In Figure \ref{fig:matching_probability}, we plot the matched fraction of each pair of ions in several of our cloud-crushing simulations at t50. The colour of each grid indicates the probability that an absorber of the ion on the $x$-axis is matched to another absorber of the ion on the $y$-axis. \revision{It is worth noting that these figures are not diagonally symmetric. For example, if all Mg \textsc{ii} absorbers have matched O \textsc{ii} absorbers in a simulation, then the colour of the pixel at the location (2, 3) will correspond to 1.0. On the other hand, there can still be a large number of ``stand-alone" O \textsc{ii} absorbers that do not match any of the Mg \textsc{ii} absorbers, in which case the colour of the pixel at the location (3, 2) will correspond to a much smaller number.} The top row shows the matching fractions of ions in three non-conductive simulations with different $M$ and $\chi_0$ values, and the bottom row shows those of their weakly-conductive counterparts. Comparing the upper left parts of all panels, which show the matched fractions of lower ions to higher ones, we note that both the non-conductive and conductive simulations have high fractions of matches between low and medium ions with ionization potentials below that of C \textsc{iv}. The matched fractions between the low ions and higher ions are larger in the non-conductive runs than in their conductive counterparts. However, we note that none of our simulations produce a large number of matching absorbers between the low ions and O \textsc{vi}.}

\yangedit{Comparing the lower right parts of the panels on the top row with those on the bottom row, we find that the high ion absorbers in the non-conductive simulation are predominantly isolated, i. e. not matched to absorbers of lower ions. On the other hand, the matching probability of absorbers of high ions to those of low ions is much higher in the conductive simulations. It is interesting that the ion-matching patterns depend much more strongly on the thermal conduction than the Mach number or the initial density contrast. The other simulations, which are not shown in Figure \ref{fig:matching_probability}, also follow very similar trends.}

\begin{figure*}
	\centering
	\includegraphics[width=0.99\linewidth]{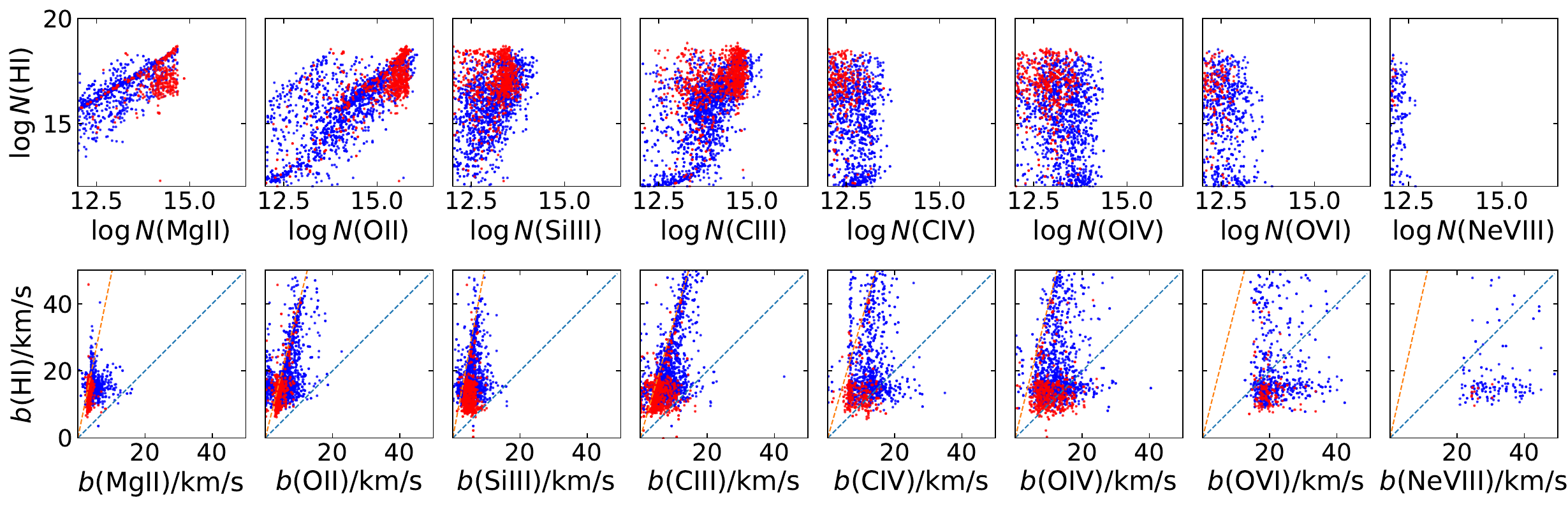}
	\caption{Co-distributions of the column densities and line widths of matching absorbers between H \textsc{i} and other ions in the T0.3\_v1700\_chi300 \revision{(blue dots)} and T0.3\_v1700\_chi300\_cond\_0.1 \revision{(red dots)} simulations at t50, $z=0.540$ and a projection angle of $\theta=60^{\circ}$. As a reference, two dashed lines are plotted on each panel, indicating absorbers with equal line widths (green lines) and thermally broadened absorbers with the same temperature (orange lines).}
	\label{fig:HI_correlation_T0.3_v1700_chi300_ld2_t2_5_z0.540}
\end{figure*}
Interesting patterns can also be seen in the column density and line width co-distributions of these matching pairs. In Figure \ref{fig:HI_correlation_T0.3_v1700_chi300_ld2_t2_5_z0.540}, we plot the co-distributions of the line widths of matching absorbers between H \textsc{i} and other ions. Here we treat pairs of absorbers matched in the order of increasing velocity offset and those matched in the reverse order as independent, separate matches. On the bottom panels, the green dashed lines indicate matched absorbers with equal line widths, and the orange dashed lines denote matched absorbers that are completely thermally broadened with the same temperature for both ions. We find that the non-conductive simulation shows a very clear high-line width streak in the co-distributions of the line widths of matched absorbers between H \textsc{i} and ions with ionization potentials higher than Mg \textsc{ii}. For ions below O \textsc{iv}, this streak clearly follows the equal-temperature orange line, meaning that these absorbers are thermally-broadened with equal temperatures for both ions. For O \textsc{iv} and O \textsc{vi}, there seems to be more kinematic broadening. While there are also a few pairs of absorbers in T0.3\_v1700\_chi300\_cond\_0.1 on this line, they occur much less frequently than those in T0.3\_v1700\_chi300. These streaks can also be seen in the column density co-distributions on the top panels, where they appear on the bottom. We attribute them to the fragments inside the mixing layer in the non-conductive simulation, which have lower densities than the core, allowing the temperatures to stay higher, but not hot enough to completely destroy these ions.

\section{Discussions and Conclusions}
\label{sec:conclusions}
In this work, we conduct a series of cloud-crushing simulations similar to those carried out by \citet{Scannapieco2015} and \citet{Brueggen2016}, but with lower cloud densities that match the CGM of a Milky Way-like galaxy. These are characterized by two dimensionless numbers: the initial density contrast, $\chi_0$, and the Mach number of the ambient flow, \revision{$M_{\mathrm{hot}}$}. In addition, we run our simulations with three different levels of conduction: no conduction, full Spitzer conduction and weak conduction, which has a reduction factor of $\eta=0.1$ from Spitzer (see Equation \ref{eq:kappa}).

For each simulation, we compute the ionization fractions of various ions and generate ion maps and mock absorption spectra from our simulations \revision{assuming ionization equilibrium in the presence of a ionizing background}. We find that different ions reside in different regions in our clouds. In particular, low ions concentrate at the center of the clouds, while higher ones are more spread out and extend to the edge of the clouds. The mock spectra generated from our simulations show the coexistence of multiple ions with vastly different ionization potentials \revision{at the same position in velocity space}. However, our models are unable to generate a significant amount of O \textsc{vi},
which is sometimes found to coexist with H \textsc{i} and other low ions in observed absorption systems.

Through fitting the mock absorption spectra, we obtain lists of absorbers and study the statistical distributions of their column densities and line widths. We find a significant impact of thermal conduction on the ion distributions in most of our simulations, the concrete form of which depends on the Mach number and the density contrast. The only exceptions are the simulations with the lowest Mach numbers and density contrasts, T0.1\_v150\_chi100 and T0.1\_v150\_chi100\_cond\_0.1, where only the high ions O \textsc{vi} and Ne \textsc{viii} are affected by thermal conduction. Tracking their evolution, we find that the ion column density distributions of the clouds in the non-conductive simulations do not significantly evolve, while those in their conductive counterparts vary more over time, in accordance with the initial evaporative compression and subsequent stretching of the clouds. In addition, we also explore the dependence of the distributions on the projection angles, and find a clear distinction between the head-on projections ($\theta=0^{\circ}$) and all other directions in both the conductive and non-conductive simulations.

By comparing the column densities of absorbers exposed to the UV ionizating backgrounds at different redshifts, as well as those in the absence of photoionization, we can determine the contribution of photoionization to each ion. In all of our simulations, the low ionization potentials ions H \textsc{i} and Mg \textsc{ii} are destroyed by photoionization, while the number of high column density absorbers with ionization potentials between those of O \textsc{ii} and C \textsc{iii} are significantly increased by photoionization.  We also find that the high ions O \textsc{vi} and Ne \textsc{viii} are almost entirely collisionally ionized and unaffected by the ionizing background.

We have emphasized the impact of various parameters on the column density and line width distributions of the ions in our clouds. However, it is also worth noting that, with the exception of the T0.3\_v3000\_chi300\_cond\_0.1 simulation, in which the morphology of the cloud is significantly different due to inefficient cooling, the shapes of the column density and line-width distributions are quite similar across all of our simulations at the same conduction level. 

We also explored the correlation between absorbers of different ions in the same absorption systems. We find that a significant fraction of low ion absorbers have matching middle ion absorbers with a velocity offset within $\Delta v=10\mathrm{km/s}$. However, as with the mock absorption lines, there is a lack of high ion absorbers that are matched with the low ions in our simulations. We also find a clear separation between absorbers in the center regions and those in the fragments of the clouds in the non-conductive simulations, indicating that the fragments have higher temperatures and lower densities.

Our results provide a physical foundation for interpreting QSO absorption observations. As discussed in \S\ref{subsec:mock_spectra}, when comparing our results with observations, the column density histograms of our absorbers can be horizontally shifted, depending on the mass column densities and metallicities of the observed clouds, as well as the ionizing background. Our clouds display the coexistence of ions with vastly different ionization potentials, both from individual absorption spectra, such as those shown in Figures \ref{fig:absorption_example} and \ref{fig:absorption_example_cond}, and the matched fractions between low and middle ions shown in Figure \ref{fig:matching_probability}. However, our simulations are unable to explain the absorption systems with aligned H \textsc{i} and O \textsc{vi} that have been widely observed.

One possible explanation is that the clouds in these observed galaxies have larger sizes and masses than our clouds, thereby requiring an upscaling of the column densities of metal ions. In addition, we have used a solar metallicity throughout our study, while in observed absorption systems, the clouds can have super-solar metallicities, which results in more O \textsc{vi}. \revision{Furthermore, we have assumed ionization equilibrium throughout our work, while nonequilibrium effects could increase the production of O \textsc{vi}}. \yangedit{Another possible factor that affects the coexistence of low and high ions is the conduction model. For example, \citet{Brueggen2023} performed a series of anisotropically conductive cloud-crushing simulations, where anisotropy in the thermal conduction was introduced through the effect of magnetic fields, and found that the morphology of the clouds were significantly altered by the anisotropic conduction. In some cases, their clouds displayed a wider range of $\rho$ and $T$, which could help in having low and high excitation ions coexist.} \revision{More broadly, magnetic fields can suppress the KHI in the clouds, reduce their mixing with the ambient gas and, depending on their orientations, either enhance or suppress the shock compression of the clouds (see, for example, \citealt{Cottle2020}), all of which can affect the ionization fractions}. Alternatively, these observed O \textsc{vi} lines may originate from the hot phase of the CGM that is somehow aligned with the clouds. It is also possible that these lines come from collections of clouds that travel together and have similar velocity offsets as a result. Even though each cloud does not contain an observable amount of O \textsc{vi}, a quasar sightline might go through multiple clouds and the combined column density can be much higher.

In future work, we will establish a routine that translates our results to cosmological simulations that use physics-driven, multiphase wind models, such as the PhEW model. This will enable us to study the ion distributions in the CGM of the galaxies in these simulations while properly accounting for the multiphase and non-uniform nature of the clouds in the outflows. Another question we find worthwhile for future studies is to address how the cooling efficiency affects the evolution of the clouds. As discussed in \S \ref{sec:mass_loss}, some of our clouds have cooling time scales that are comparable with their cloud crushing times both in their mixing layers and in their cores, and the evolution of the clouds is affected by the inefficient cooling in these cases. A systematic exploration of this will help establish a criterion for the conditions under which cool clouds are able to survive in the CGM and provide insights on their physical properties. 

\section*{Acknowledgements}

\yangedit{We would like to acknowledge J'Neil Cottle and Todd Tripp for helpful discussions that greatly improved this manuscript. NSK, ES and LY acknowledge support by NASA grant 80NSSC25K7299. NSK and LY acknowledge support by NASA ATP grant 80NSSC18K1016. NSK also acknowledges funding from NSF grant AST-2205725. ES acknowledges support by NASA grant 80NSSC25PA052.  MB acknowledges support from the Deutsche Forschungsgemeinschaft under Germany's Excellence Strategy - EXC 2121 ``Quantum Universe" - 390833306 and from the BMBF ErUM-Pro grant 05A2023. The work includes simulations run on NASA High-End Computing resources. The analyses of our simulations were performed on the Unity cluster at the Massachusetts Green High Performance Computing Center (MGHPCC).}

\section*{Data availability}
The data underlying this article will be shared on reasonable request to the corresponding author.

\bibliographystyle{mnras}
\bibliography{quasar_absorption}

\begin{thebibliography}{}
\makeatletter
\relax
\def\mn@urlcharsother{\let\do\@makeother \do\$\do\&\do\#\do\^\do\_\do\%\do\~}
\def\mn@doi{\begingroup\mn@urlcharsother \@ifnextchar [ {\mn@doi@}
  {\mn@doi@[]}}
\def\mn@doi@[#1]#2{\def\@tempa{#1}\ifx\@tempa\@empty \href
  {http://dx.doi.org/#2} {doi:#2}\else \href {http://dx.doi.org/#2} {#1}\fi
  \endgroup}
\def\mn@eprint#1#2{\mn@eprint@#1:#2::\@nil}
\def\mn@eprint@arXiv#1{\href {http://arxiv.org/abs/#1} {{\tt arXiv:#1}}}
\def\mn@eprint@dblp#1{\href {http://dblp.uni-trier.de/rec/bibtex/#1.xml}
  {dblp:#1}}
\def\mn@eprint@#1:#2:#3:#4\@nil{\def\@tempa {#1}\def\@tempb {#2}\def\@tempc
  {#3}\ifx \@tempc \@empty \let \@tempc \@tempb \let \@tempb \@tempa \fi \ifx
  \@tempb \@empty \def\@tempb {arXiv}\fi \@ifundefined
  {mn@eprint@\@tempb}{\@tempb:\@tempc}{\expandafter \expandafter \csname
  mn@eprint@\@tempb\endcsname \expandafter{\@tempc}}}

\bibitem[\protect\citeauthoryear{{Anderson}, {Gaspari}, {White}, {Wang}  \&
  {Dai}}{{Anderson} et~al.}{2015}]{Anderson2015}
{Anderson} M.~E.,  {Gaspari} M.,  {White} S. D.~M.,  {Wang} W.,   {Dai} X.,
  2015, \mn@doi [\mnras] {10.1093/mnras/stv437}, \href
  {https://ui.adsabs.harvard.edu/abs/2015MNRAS.449.3806A} {449, 3806}

\bibitem[\protect\citeauthoryear{Appleby, Dav{\'{e}}, Sorini, Storey-Fisher  \&
  Smith}{Appleby et~al.}{2021}]{Appleby2021}
Appleby S.,  Dav{\'{e}} R.,  Sorini D.,  Storey-Fisher K.,   Smith B.,  2021,
  \mn@doi [Monthly Notices of the Royal Astronomical Society]
  {10.1093/mnras/stab2310}, 507, 2383

\bibitem[\protect\citeauthoryear{Appleby, Davé, Sorini, Cui  \&
  Christiansen}{Appleby et~al.}{2022}]{Appleby2022}
Appleby S.,  Davé R.,  Sorini D.,  Cui W.,   Christiansen J.,  2022, The
  Physical Nature of Circumgalactic Medium Absorbers in Simba,
  \mn@doi{10.48550/ARXIV.2207.04068}, \url {https://arxiv.org/abs/2207.04068}

\bibitem[\protect\citeauthoryear{Banda-Barrag{\'{a}}n, Brüggen, Federrath,
  Wagner, Scannapieco  \& Cottle}{Banda-Barrag{\'{a}}n
  et~al.}{2020}]{BandaBarragan2020}
Banda-Barrag{\'{a}}n W.~E.,  Brüggen M.,  Federrath C.,  Wagner A.~Y.,
  Scannapieco E.,   Cottle J.,  2020, \mn@doi [Monthly Notices of the Royal
  Astronomical Society] {10.1093/mnras/staa2904}, 499, 2173

\bibitem[\protect\citeauthoryear{Banda-Barrag{\'{a}}n, Brüggen, Heesen,
  Scannapieco, Cottle, Federrath  \& Wagner}{Banda-Barrag{\'{a}}n
  et~al.}{2021}]{BandaBarragan2021}
Banda-Barrag{\'{a}}n W.~E.,  Brüggen M.,  Heesen V.,  Scannapieco E.,  Cottle
  J.,  Federrath C.,   Wagner A.~Y.,  2021, \mn@doi [Monthly Notices of the
  Royal Astronomical Society] {10.1093/mnras/stab1884}, 506, 5658

\bibitem[\protect\citeauthoryear{Banda-Barragán, Zertuche, Federrath,
  García Del Valle, Brüggen  \& Wagner}{Banda-Barragán
  et~al.}{2019}]{BandaBarragan2019}
Banda-Barragán W.~E.,  Zertuche F.~J.,  Federrath C.,  García Del Valle J.,
   Brüggen M.,   Wagner A.~Y.,  2019, Mon Not R Astron Soc, 486, 4526

\bibitem[\protect\citeauthoryear{Baur, Palanque-Delabrouille, Yèche,
  Magneville  \& Viel}{Baur et~al.}{2016}]{Baur2016}
Baur J.,  Palanque-Delabrouille N.,  Yèche C.,  Magneville C.,   Viel M.,
  2016, \mn@doi [Journal of Cosmology and Astroparticle Physics]
  {10.1088/1475-7516/2016/08/012}, 2016, 012

\bibitem[\protect\citeauthoryear{{Bechtold}}{{Bechtold}}{2001}]{Bechtold2001}
{Bechtold} J.,  2001, \mn@doi [arXiv e-prints]
  {10.48550/arXiv.astro-ph/0112521}, \href
  {https://ui.adsabs.harvard.edu/abs/2001astro.ph.12521B} {pp
  astro--ph/0112521}

\bibitem[\protect\citeauthoryear{Birkinshaw}{Birkinshaw}{1999}]{Birkinshaw1999}
Birkinshaw M.,  1999, Physics Reports, 310, 97

\bibitem[\protect\citeauthoryear{Bordoloi, Prochaska, Tumlinson, Werk, Tripp
  \& Burchett}{Bordoloi et~al.}{2018}]{Bordoloi2018}
Bordoloi R.,  Prochaska J.~X.,  Tumlinson J.,  Werk J.~K.,  Tripp T.~M.,
  Burchett J.~N.,  2018, \mn@doi [The Astrophysical Journal]
  {10.3847/1538-4357/aad8ac}, 864, 132

\bibitem[\protect\citeauthoryear{Brüggen \& Scannapieco}{Brüggen \&
  Scannapieco}{2016}]{Brueggen2016}
Brüggen M.,  Scannapieco E.,  2016, The Astrophysical Journal, 822, 31

\bibitem[\protect\citeauthoryear{Brüggen, Scannapieco  \& Grete}{Brüggen
  et~al.}{2023}]{Brueggen2023}
Brüggen M.,  Scannapieco E.,   Grete P.,  2023, \mn@doi [The Astrophysical
  Journal] {10.3847/1538-4357/acd63e}, 951, 113

\bibitem[\protect\citeauthoryear{Casavecchia, Banda-Barragán, Brüggen,
  Brighenti  \& Scannapieco}{Casavecchia et~al.}{2024}]{Casavecchia2024}
Casavecchia B.,  Banda-Barragán W.~E.,  Brüggen M.,  Brighenti F.,
  Scannapieco E.,  2024, \mn@doi [Astronomy &amp; Astrophysics]
  {10.1051/0004-6361/202449461}, 689, A127

\bibitem[\protect\citeauthoryear{{Chatzikos} et~al.,}{{Chatzikos}
  et~al.}{2023}]{Chatzikos2023}
{Chatzikos} M.,  et~al., 2023, \mn@doi [\rmxaa]
  {10.22201/ia.01851101p.2023.59.02.12}, \href
  {https://ui.adsabs.harvard.edu/abs/2023RMxAA..59..327C} {59, 327}

\bibitem[\protect\citeauthoryear{Chen, Wild, Tinker, Gauthier, Helsby, Shectman
   \& Thompson}{Chen et~al.}{2010}]{Chen2010}
Chen H.-W.,  Wild V.,  Tinker J.~L.,  Gauthier J.-R.,  Helsby J.~E.,  Shectman
  S.~A.,   Thompson I.~B.,  2010, \mn@doi [The Astrophysical Journal]
  {10.1088/2041-8205/724/2/l176}, 724, L176

\bibitem[\protect\citeauthoryear{Chevalier \& Clegg}{Chevalier \&
  Clegg}{1985}]{Chevalier1985}
Chevalier R.~A.,  Clegg A.~W.,  1985, Nature, 317, 44

\bibitem[\protect\citeauthoryear{Chisholm, Tremonti, Leitherer, Chen  \&
  Wofford}{Chisholm et~al.}{2016}]{Chisholm2016}
Chisholm J.,  Tremonti C.~A.,  Leitherer C.,  Chen Y.,   Wofford A.,  2016,
  \mn@doi [Monthly Notices of the Royal Astronomical Society]
  {10.1093/mnras/stw178}, 457, 3133

\bibitem[\protect\citeauthoryear{Chisholm, Bordoloi, Rigby  \&
  Bayliss}{Chisholm et~al.}{2018}]{Chisholm2018}
Chisholm J.,  Bordoloi R.,  Rigby J.~R.,   Bayliss M.,  2018, Mon Not R Astron
  Soc, 474, 1688

\bibitem[\protect\citeauthoryear{Churchill, Nielsen, Kacprzak  \&
  Trujillo-Gomez}{Churchill et~al.}{2013}]{Churchill2013}
Churchill C.~W.,  Nielsen N.~M.,  Kacprzak G.~G.,   Trujillo-Gomez S.,  2013,
  \mn@doi [The Astrophysical Journal] {10.1088/2041-8205/763/2/l42}, 763, L42

\bibitem[\protect\citeauthoryear{{Cottle}, {Scannapieco}  \&
  {Br{\"u}ggen}}{{Cottle} et~al.}{2018}]{Cottle2018}
{Cottle} J.~N.,  {Scannapieco} E.,   {Br{\"u}ggen} M.,  2018, \mn@doi [\apj]
  {10.3847/1538-4357/aad55c}, \href
  {https://ui.adsabs.harvard.edu/abs/2018ApJ...864...96C} {864, 96}

\bibitem[\protect\citeauthoryear{Cottle, Scannapieco, Brüggen,
  Banda-Barrag{\'{a}}n  \& Federrath}{Cottle et~al.}{2020}]{Cottle2020}
Cottle J.,  Scannapieco E.,  Brüggen M.,  Banda-Barrag{\'{a}}n W.,   Federrath
  C.,  2020, \mn@doi [The Astrophysical Journal] {10.3847/1538-4357/ab76d1},
  892, 59

\bibitem[\protect\citeauthoryear{{Danforth} \& {Shull}}{{Danforth} \&
  {Shull}}{2008}]{Danforth2008}
{Danforth} C.~W.,  {Shull} J.~M.,  2008, \mn@doi [\apj] {10.1086/587127}, \href
  {https://ui.adsabs.harvard.edu/abs/2008ApJ...679..194D} {679, 194}

\bibitem[\protect\citeauthoryear{{Danforth} et~al.,}{{Danforth}
  et~al.}{2016}]{Danforth2016}
{Danforth} C.~W.,  et~al., 2016, \mn@doi [\apj] {10.3847/0004-637X/817/2/111},
  \href {https://ui.adsabs.harvard.edu/abs/2016ApJ...817..111D} {817, 111}

\bibitem[\protect\citeauthoryear{Davé, Oppenheimer, Katz, Kollmeier  \&
  Weinberg}{Davé et~al.}{2010}]{Dave2010}
Davé R.,  Oppenheimer B.~D.,  Katz N.,  Kollmeier J.~A.,   Weinberg D.~H.,
  2010, \mn@doi [Mon Not R Astron Soc] {10.1111/j.1365-2966.2010.17279.x}, 408,
  2051

\bibitem[\protect\citeauthoryear{{Di Matteo}, {Springel}  \& {Hernquist}}{{Di
  Matteo} et~al.}{2005}]{DiMatteo2005}
{Di Matteo} T.,  {Springel} V.,   {Hernquist} L.,  2005, \mn@doi [\nat]
  {10.1038/nature03335}, \href
  {https://ui.adsabs.harvard.edu/abs/2005Natur.433..604D} {433, 604}

\bibitem[\protect\citeauthoryear{{Fabian}}{{Fabian}}{2012}]{Fabian2012}
{Fabian} A.~C.,  2012, \mn@doi [\araa] {10.1146/annurev-astro-081811-125521},
  \href {https://ui.adsabs.harvard.edu/abs/2012ARA&A..50..455F} {50, 455}

\bibitem[\protect\citeauthoryear{Faerman, Pandya, Somerville  \&
  Sternberg}{Faerman et~al.}{2022}]{Faerman2022}
Faerman Y.,  Pandya V.,  Somerville R.~S.,   Sternberg A.,  2022, \mn@doi [The
  Astrophysical Journal] {10.3847/1538-4357/ac4ca6}, 928, 37

\bibitem[\protect\citeauthoryear{Ferland, Korista, Verner, Ferguson, Kingdon
  \& Verner}{Ferland et~al.}{1998}]{Ferland1998}
Ferland G.,  Korista K.,  Verner D.,  Ferguson J.,  Kingdon J.,   Verner E.,
  1998, \mn@doi [Publications of the Astronomical Society of the Pacific]
  {10.1086/316190}, 110, 761

\bibitem[\protect\citeauthoryear{{Ford}, {Oppenheimer}, {Dav{\'e}}, {Katz},
  {Kollmeier}  \& {Weinberg}}{{Ford} et~al.}{2013}]{Ford2013}
{Ford} A.~B.,  {Oppenheimer} B.~D.,  {Dav{\'e}} R.,  {Katz} N.,  {Kollmeier}
  J.~A.,   {Weinberg} D.~H.,  2013, \mn@doi [\mnras] {10.1093/mnras/stt393},
  \href {https://ui.adsabs.harvard.edu/abs/2013MNRAS.432...89F} {432, 89}

\bibitem[\protect\citeauthoryear{{Ford}, {Dav{\'e}}, {Oppenheimer}, {Katz},
  {Kollmeier}, {Thompson}  \& {Weinberg}}{{Ford} et~al.}{2014}]{Ford2014}
{Ford} A.~B.,  {Dav{\'e}} R.,  {Oppenheimer} B.~D.,  {Katz} N.,  {Kollmeier}
  J.~A.,  {Thompson} R.,   {Weinberg} D.~H.,  2014, \mn@doi [\mnras]
  {10.1093/mnras/stu1418}, \href
  {https://ui.adsabs.harvard.edu/abs/2014MNRAS.444.1260F} {444, 1260}

\bibitem[\protect\citeauthoryear{{Ford} et~al.,}{{Ford}
  et~al.}{2016}]{Ford2016}
{Ford} A.~B.,  et~al., 2016, \mn@doi [\mnras] {10.1093/mnras/stw595}, \href
  {https://ui.adsabs.harvard.edu/abs/2016MNRAS.459.1745F} {459, 1745}

\bibitem[\protect\citeauthoryear{Fryxell et~al.,}{Fryxell
  et~al.}{2000}]{Fryxell2000}
Fryxell B.,  et~al., 2000, The Astrophysical Journal Supplement Series, 131,
  273

\bibitem[\protect\citeauthoryear{Haardt \& Madau}{Haardt \&
  Madau}{2012}]{Haardt2012}
Haardt F.,  Madau P.,  2012, The Astrophysical Journal, 746, 125

\bibitem[\protect\citeauthoryear{Haislmaier, Tripp, Katz, Prochaska, Burchett,
  O'Meara  \& Werk}{Haislmaier et~al.}{2020}]{Haislmaier2020}
Haislmaier K.~J.,  Tripp T.~M.,  Katz N.,  Prochaska J.~X.,  Burchett J.~N.,
  O'Meara J.~M.,   Werk J.~K.,  2020, \mn@doi [Monthly Notices of the Royal
  Astronomical Society] {10.1093/mnras/staa3544}, 502, 4993

\bibitem[\protect\citeauthoryear{{Heckman}, {Armus}  \& {Miley}}{{Heckman}
  et~al.}{1990}]{Heckman1990}
{Heckman} T.~M.,  {Armus} L.,   {Miley} G.~K.,  1990, \mn@doi [\apjs]
  {10.1086/191522}, \href
  {https://ui.adsabs.harvard.edu/abs/1990ApJS...74..833H} {74, 833}

\bibitem[\protect\citeauthoryear{{Hernquist}, {Katz}, {Weinberg}  \&
  {Miralda-Escud{\'e}}}{{Hernquist} et~al.}{1996}]{Hernquist1996}
{Hernquist} L.,  {Katz} N.,  {Weinberg} D.~H.,   {Miralda-Escud{\'e}} J.,
  1996, \mn@doi [\apjl] {10.1086/309899}, \href
  {https://ui.adsabs.harvard.edu/abs/1996ApJ...457L..51H} {457, L51}

\bibitem[\protect\citeauthoryear{Huang, Katz, Scannapieco, Cottle, Davé,
  Weinberg, Peeples  \& Brüggen}{Huang et~al.}{2020}]{Huang2020a}
Huang S.,  Katz N.,  Scannapieco E.,  Cottle J.,  Davé R.,  Weinberg D.~H.,
  Peeples M.~S.,   Brüggen M.,  2020, Monthly Notices of the Royal
  Astronomical Society, pp 2586--2604

\bibitem[\protect\citeauthoryear{{Huang}, {Katz}, {Cottle}, {Scannapieco},
  {Dav{\'e}}  \& {Weinberg}}{{Huang} et~al.}{2021}]{Huang2021}
{Huang} S.,  {Katz} N.,  {Cottle} J.,  {Scannapieco} E.,  {Dav{\'e}} R.,
  {Weinberg} D.~H.,  2021, arXiv e-prints, \href
  {https://ui.adsabs.harvard.edu/abs/2021arXiv210601511H} {p. arXiv:2106.01511}

\bibitem[\protect\citeauthoryear{Hummels, Bryan, Smith  \& Turk}{Hummels
  et~al.}{2013}]{Hummels2013}
Hummels C.~B.,  Bryan G.~L.,  Smith B.~D.,   Turk M.~J.,  2013, \mn@doi
  [Monthly Notices of the Royal Astronomical Society] {10.1093/mnras/sts702},
  430, 1548

\bibitem[\protect\citeauthoryear{Hummels, Smith  \& Silvia}{Hummels
  et~al.}{2017}]{Hummels2017}
Hummels C.~B.,  Smith B.~D.,   Silvia D.~W.,  2017, The Astrophysical Journal,
  847, 59

\bibitem[\protect\citeauthoryear{{Johnson}, {Chen}, {Mulchaey}, {Schaye}  \&
  {Straka}}{{Johnson} et~al.}{2017}]{Johnson2017}
{Johnson} S.~D.,  {Chen} H.-W.,  {Mulchaey} J.~S.,  {Schaye} J.,   {Straka}
  L.~A.,  2017, \mn@doi [\apjl] {10.3847/2041-8213/aa9370}, \href
  {https://ui.adsabs.harvard.edu/abs/2017ApJ...850L..10J} {850, L10}

\bibitem[\protect\citeauthoryear{Kauffmann, Nelson, Borthakur, Heckman,
  Hernquist, Marinacci, Pakmor  \& Pillepich}{Kauffmann
  et~al.}{2019}]{Kauffmann2019}
Kauffmann G.,  Nelson D.,  Borthakur S.,  Heckman T.,  Hernquist L.,  Marinacci
  F.,  Pakmor R.,   Pillepich A.,  2019, \mn@doi [Monthly Notices of the Royal
  Astronomical Society] {10.1093/mnras/stz1029}, 486, 4686

\bibitem[\protect\citeauthoryear{{King} \& {Pounds}}{{King} \&
  {Pounds}}{2015}]{King2015}
{King} A.,  {Pounds} K.,  2015, \mn@doi [\araa]
  {10.1146/annurev-astro-082214-122316}, \href
  {https://ui.adsabs.harvard.edu/abs/2015ARA&A..53..115K} {53, 115}

\bibitem[\protect\citeauthoryear{{Kirkman}, {Tytler}, {Lubin}  \&
  {Charlton}}{{Kirkman} et~al.}{2007}]{Kirkman2007}
{Kirkman} D.,  {Tytler} D.,  {Lubin} D.,   {Charlton} J.,  2007, \mn@doi
  [\mnras] {10.1111/j.1365-2966.2007.11502.x}, \href
  {https://ui.adsabs.harvard.edu/abs/2007MNRAS.376.1227K} {376, 1227}

\bibitem[\protect\citeauthoryear{{Klein}, {McKee}  \& {Colella}}{{Klein}
  et~al.}{1994}]{Klein1994}
{Klein} R.~I.,  {McKee} C.~F.,   {Colella} P.,  1994, \mn@doi [\apj]
  {10.1086/173554}, \href
  {https://ui.adsabs.harvard.edu/abs/1994ApJ...420..213K} {420, 213}

\bibitem[\protect\citeauthoryear{Lim, Barnes, Vogelsberger, Mo, Nelson,
  Pillepich, Dolag  \& Marinacci}{Lim et~al.}{2020}]{Lim2020}
Lim S.~H.,  Barnes D.,  Vogelsberger M.,  Mo H.~J.,  Nelson D.,  Pillepich A.,
  Dolag K.,   Marinacci F.,  2020, \mn@doi [Monthly Notices of the Royal
  Astronomical Society] {10.1093/mnras/stab1172}, 504, 5131

\bibitem[\protect\citeauthoryear{Lochhaas et~al.,}{Lochhaas
  et~al.}{2023}]{Lochhaas2023}
Lochhaas C.,  et~al., 2023, \mn@doi [The Astrophysical Journal]
  {10.3847/1538-4357/acbb06}, 948, 43

\bibitem[\protect\citeauthoryear{McCourt, Oh, O'Leary  \& Madigan}{McCourt
  et~al.}{2018}]{McCourt2018}
McCourt M.,  Oh S.~P.,  O'Leary R.,   Madigan A.-M.,  2018, Mon Not R Astron
  Soc, 473, 5407

\bibitem[\protect\citeauthoryear{Meiring, Tripp, Werk, Howk, Jenkins,
  Prochaska, Lehner  \& Sembach}{Meiring et~al.}{2013}]{Meiring2013}
Meiring J.~D.,  Tripp T.~M.,  Werk J.~K.,  Howk J.~C.,  Jenkins E.~B.,
  Prochaska J.~X.,  Lehner N.,   Sembach K.~R.,  2013, \mn@doi [The
  Astrophysical Journal] {10.1088/0004-637x/767/1/49}, 767, 49

\bibitem[\protect\citeauthoryear{Mo, Van~den Bosch  \& White}{Mo
  et~al.}{2010}]{Mo2010}
Mo H.,  Van~den Bosch F.,   White S.,  2010, Galaxy formation and evolution.
Cambridge University Press

\bibitem[\protect\citeauthoryear{Oort}{Oort}{1981}]{Oort1981}
Oort J.,  1981, Astronomy and Astrophysics, Vol. 94, P. 359, 1981, 94, 359

\bibitem[\protect\citeauthoryear{Oppenheimer \& Davé}{Oppenheimer \&
  Davé}{2006}]{Oppenheimer2006}
Oppenheimer B.~D.,  Davé R.,  2006, Mon Not R Astron Soc, 373, 1265

\bibitem[\protect\citeauthoryear{Oppenheimer, Schaye, Crain, Werk  \&
  Richings}{Oppenheimer et~al.}{2018}]{Oppenheimer2018}
Oppenheimer B.~D.,  Schaye J.,  Crain R.~A.,  Werk J.~K.,   Richings A.~J.,
  2018, \mn@doi [Monthly Notices of the Royal Astronomical Society]
  {10.1093/mnras/sty2281}, 481, 835

\bibitem[\protect\citeauthoryear{{Orlando}, {Peres}, {Reale}, {Bocchino},
  {Rosner}, {Plewa}  \& {Siegel}}{{Orlando} et~al.}{2005}]{Orlando2005}
{Orlando} S.,  {Peres} G.,  {Reale} F.,  {Bocchino} F.,  {Rosner} R.,  {Plewa}
  T.,   {Siegel} A.,  2005, \mn@doi [\aap] {10.1051/0004-6361:20052896}, \href
  {https://ui.adsabs.harvard.edu/abs/2005A&A...444..505O} {444, 505}

\bibitem[\protect\citeauthoryear{Putman, Peek  \& Joung}{Putman
  et~al.}{2012}]{Putman2012}
Putman M.~E.,  Peek J. E.~G.,   Joung M.~R.,  2012, Annual Review of Astronomy
  and Astrophysics, pp 491--529

\bibitem[\protect\citeauthoryear{Reed, Gardner, Quinn, Stadel, Fardal, Lake  \&
  Governato}{Reed et~al.}{2003}]{Reed2003}
Reed D.,  Gardner J.,  Quinn T.,  Stadel J.,  Fardal M.,  Lake G.,   Governato
  F.,  2003, \mn@doi [Monthly Notices of the Royal Astronomical Society]
  {10.1046/j.1365-2966.2003.07113.x}, 346, 565

\bibitem[\protect\citeauthoryear{Rudie, Steidel, Pettini, Trainor, Strom,
  Hummels, Reddy  \& Shapley}{Rudie et~al.}{2019}]{Rudie2019}
Rudie G.~C.,  Steidel C.~C.,  Pettini M.,  Trainor R.~F.,  Strom A.~L.,
  Hummels C.~B.,  Reddy N.~A.,   Shapley A.~E.,  2019, \mn@doi [The
  Astrophysical Journal] {10.3847/1538-4357/ab4255}, 885, 61

\bibitem[\protect\citeauthoryear{Scannapieco \& Brüggen}{Scannapieco \&
  Brüggen}{2015}]{Scannapieco2015}
Scannapieco E.,  Brüggen M.,  2015, The Astrophysical Journal, 805, 158

\bibitem[\protect\citeauthoryear{Scannapieco \& Oh}{Scannapieco \&
  Oh}{2004}]{Scannapieco2004}
Scannapieco E.,  Oh S.~P.,  2004, \mn@doi [The Astrophysical Journal]
  {10.1086/386542}, 608, 62

\bibitem[\protect\citeauthoryear{Schneider \& Robertson}{Schneider \&
  Robertson}{2017}]{Schneider2017}
Schneider E.~E.,  Robertson B.~E.,  2017, \mn@doi [The Astrophysical Journal]
  {10.3847/1538-4357/834/2/144}, 834, 144

\bibitem[\protect\citeauthoryear{Seljak, Slosar  \& McDonald}{Seljak
  et~al.}{2006}]{Seljak2006}
Seljak U.,  Slosar A.,   McDonald P.,  2006, \mn@doi [Journal of Cosmology and
  Astroparticle Physics] {10.1088/1475-7516/2006/10/014}, 2006, 014

\bibitem[\protect\citeauthoryear{{Stocke}, {Keeney}, {Danforth}, {Shull},
  {Froning}, {Green}, {Penton}  \& {Savage}}{{Stocke}
  et~al.}{2013}]{Stocke2013}
{Stocke} J.~T.,  {Keeney} B.~A.,  {Danforth} C.~W.,  {Shull} J.~M.,  {Froning}
  C.~S.,  {Green} J.~C.,  {Penton} S.~V.,   {Savage} B.~D.,  2013, \mn@doi
  [\apj] {10.1088/0004-637X/763/2/148}, \href
  {https://ui.adsabs.harvard.edu/abs/2013ApJ...763..148S} {763, 148}

\bibitem[\protect\citeauthoryear{Strawn, Roca-Fàbrega  \& Primack}{Strawn
  et~al.}{2022}]{Strawn2022}
Strawn C.,  Roca-Fàbrega S.,   Primack J.,  2022, \mn@doi [Monthly Notices of
  the Royal Astronomical Society] {10.1093/mnras/stac3567}, 519, 1

\bibitem[\protect\citeauthoryear{Tremonti et~al.,}{Tremonti
  et~al.}{2004}]{Tremonti2004}
Tremonti C.~A.,  et~al., 2004, \mn@doi [The Astrophysical Journal]
  {10.1086/423264}, 613, 898

\bibitem[\protect\citeauthoryear{{Tripp}, {Sembach}, {Bowen}, {Savage},
  {Jenkins}, {Lehner}  \& {Richter}}{{Tripp} et~al.}{2008}]{Tripp2008}
{Tripp} T.~M.,  {Sembach} K.~R.,  {Bowen} D.~V.,  {Savage} B.~D.,  {Jenkins}
  E.~B.,  {Lehner} N.,   {Richter} P.,  2008, \mn@doi [\apjs] {10.1086/587486},
  \href {https://ui.adsabs.harvard.edu/abs/2008ApJS..177...39T} {177, 39}

\bibitem[\protect\citeauthoryear{{Tripp} et~al.,}{{Tripp}
  et~al.}{2011}]{Tripp2011}
{Tripp} T.~M.,  et~al., 2011, \mn@doi [Science] {10.1126/science.1209850},
  \href {https://ui.adsabs.harvard.edu/abs/2011Sci...334..952T} {334, 952}

\bibitem[\protect\citeauthoryear{Tumlinson et~al.,}{Tumlinson
  et~al.}{2011}]{Tumlinson2011}
Tumlinson J.,  et~al., 2011, Science, pp 948--952

\bibitem[\protect\citeauthoryear{Tumlinson, Peeples  \& Werk}{Tumlinson
  et~al.}{2017}]{Tumlinson2017}
Tumlinson J.,  Peeples M.~S.,   Werk J.~K.,  2017, Annual Review of Astronomy
  and Astrophysics, pp 389--432

\bibitem[\protect\citeauthoryear{{Turk}, {Smith}, {Oishi}, {Skory}, {Skillman},
  {Abel}  \& {Norman}}{{Turk} et~al.}{2011}]{Turk2011}
{Turk} M.~J.,  {Smith} B.~D.,  {Oishi} J.~S.,  {Skory} S.,  {Skillman} S.~W.,
  {Abel} T.,   {Norman} M.~L.,  2011, \mn@doi [\apjs]
  {10.1088/0067-0049/192/1/9}, \href
  {https://ui.adsabs.harvard.edu/abs/2011ApJS..192....9T} {192, 9}

\bibitem[\protect\citeauthoryear{{Wang} et~al.,}{{Wang}
  et~al.}{2014}]{Wang2014}
{Wang} L.,  et~al., 2014, \mn@doi [\mnras] {10.1093/mnras/stt2481}, \href
  {https://ui.adsabs.harvard.edu/abs/2014MNRAS.439..611W} {439, 611}

\bibitem[\protect\citeauthoryear{Weinberg}{Weinberg}{2003}]{Weinberg2003}
Weinberg D.~H.,  2003, in {AIP} Conference Proceedings. {AIP},
  \mn@doi{10.1063/1.1581786}

\bibitem[\protect\citeauthoryear{Werk et~al.,}{Werk et~al.}{2014}]{Werk2014}
Werk J.~K.,  et~al., 2014, \mn@doi [The Astrophysical Journal]
  {10.1088/0004-637x/792/1/8}, 792, 8

\bibitem[\protect\citeauthoryear{{Wiersma}, {Schaye}  \& {Smith}}{{Wiersma}
  et~al.}{2009}]{Wiersma2009}
{Wiersma} R. P.~C.,  {Schaye} J.,   {Smith} B.~D.,  2009, \mn@doi [\mnras]
  {10.1111/j.1365-2966.2008.14191.x}, \href
  {https://ui.adsabs.harvard.edu/abs/2009MNRAS.393...99W} {393, 99}

\bibitem[\protect\citeauthoryear{{Xu} \& {Stone}}{{Xu} \&
  {Stone}}{1995}]{Xu1995}
{Xu} J.,  {Stone} J.~M.,  1995, \mn@doi [\apj] {10.1086/176475}, \href
  {https://ui.adsabs.harvard.edu/abs/1995ApJ...454..172X} {454, 172}

\bibitem[\protect\citeauthoryear{Zahedy et~al.,}{Zahedy
  et~al.}{2021}]{Zahedy2021}
Zahedy F.~S.,  et~al., 2021, \mn@doi [Monthly Notices of the Royal Astronomical
  Society] {10.1093/mnras/stab1661}, 506, 877

\bibitem[\protect\citeauthoryear{de~la Cruz, Schneider  \& Ostriker}{de~la Cruz
  et~al.}{2021}]{Cruz2021}
de~la Cruz L.~M.,  Schneider E.~E.,   Ostriker E.~C.,  2021, \mn@doi [The
  Astrophysical Journal] {10.3847/1538-4357/ac04ac}, 919, 112

\makeatother
\end{thebibliography}

\appendix
\section{Density of a cloud in the CGM}
\label{app:cloud_density}
Using a halo overdensity $\Delta_{\mathrm{h}}=200/\Omega_{\mathrm{m}}$,
the radius of a halo is given by 
\begin{equation*}
r_{\mathrm{h}}=\text{\ensuremath{\left(\frac{GM_{\mathrm{h}}}{100H^{2}}\right)}}^{1/3}=206.3\ \mathrm{kpc}\cdot\text{\ensuremath{\left[\frac{M_{\mathrm{h},12}}{\Omega_{\mathrm{m}}(1+z)^{3}+\Omega_{\Lambda}}\right]}}^{1/3},
\end{equation*}
where $M_{\mathrm{h},12}=M_{\mathrm{h}}/10^{12}M_{\odot}$ and we adopt a Hubble
parameter of $h=0.7$. Assuming the CGM follows an inverse
square law distribution truncated at $r_{\mathrm{h}}$, the CGM density
profile is then
\begin{align*}
&\rho_{\mathrm{CGM}}=\frac{M_{\mathrm{CGM}}}{4\pi r_{\mathrm{h}}r^{2}}\\
&=6.134\times10^{-29}\ \mathrm{g/cm^{3}}\cdot M_{\mathrm{CGM},11}\cdot x^{-2}\cdot\frac{\Omega_{\mathrm{m}}(1+z)^{3}+\Omega_{\Lambda}}{M_{\mathrm{h},12}},
\end{align*}
where $M_{\mathrm{CGM},11}=M_{\mathrm{CGM}}/10^{11}M_{\odot}$ and
$x=r/r_{\mathrm{h}}$. From equation (8.45) of \citet{Mo2010}, the
virial temperature of a truncated isothermal sphere equals
\begin{align*}
T_{\mathrm{vir}}&=3.6\times10^{5}\ \mathrm{K}\cdot\left(\frac{v_{\mathrm{c}}}{100\ \mathrm{km/s}}\right)^{2}\\
&=7.51\times10^{5}\ \mathrm{K}\cdot M_{\mathrm{h},12}^{2/3}\cdot\left[\Omega_{\mathrm{m}}(1+z)^{3}+\Omega_{\Lambda}\right]^{1/3}.
\end{align*}
Assuming that both the CGM and the cloud are fully ionized and in thermal pressure equilibrium, we therefore have, for a cloud at a temperature of $10^4\ \mathrm{K}$,
\begin{align}
	&\rho_{\mathrm{c}}=\frac{\rho_{\mathrm{CGM}}T_{\mathrm{vir}}}{T_{\mathrm{c}}}\nonumber\\
    &= 4.61\times10^{-27}\ \mathrm{g/cm^3}\cdot M_{\mathrm{CGM},11}\cdot M_{\mathrm{h},12}^{-1/3}\cdot x^{-2}\cdot\left[\Omega_{\mathrm{m}}(1+z)^{3}+\Omega_{\Lambda}\right]^{4/3}.\label{eq:pressure_ratio}
\end{align}
According to \citet{Faerman2022}, the Milky Way, with a halo mass
of $\sim10^{12}\ M_{\odot}$, has a nominal hot CGM mass of $\approx5\times10^{10}\ M_{\odot}$.
Substituting these numbers, the density of a cloud at $0.5r_{\mathrm{h}}$
and $z=0$ is $9.2\times10^{-27}\ \mathrm{g/cm^3}$. For a $5\times10^{11}\ M_{\odot}$
halo, if we assume that the CGM mass is half the Milky Way value,
the density of a cloud at $0.5r_{\mathrm{h}}$ equals $5.81\times10^{-27}\ \mathrm{g/cm^3}$.
The redshift dependence is harder to obtain, since the halo mass
itself changes with time. From Figure 1 of \citet{Reed2003}, at $z=0\sim4$
the characteristic mass in the Press-Schechter formalism, $M_{*}$,
roughly scales with $z$ as $M_{*}\propto10^{-z}$. Assuming a fixed
CGM to halo mass ratio, an additional redshift-dependent factor of
$10^{-2/3\cdot z}$ needs to multiply to Equation \ref{eq:pressure_ratio}
for abundance matching. We tabulate the full redshift dependence
at a few different redshifts in Table \ref{tab:redshifts}.
\begin{center}
	\begin{table}
		\begin{centering}
			\begin{tabular}{|c|c|}
				\hline 
				$z$ & $\left[\Omega_{\mathrm{m}}(1+z)^{3}+\Omega_{\Lambda}\right]^{4/3}\cdot10^{-2/3\cdot z}$\tabularnewline
				\hline 
				0 & 1\tabularnewline
				0.5 & 0.95\tabularnewline
				1 & 0.97\tabularnewline
				2 & 0.84\tabularnewline
				4 & 0.27\tabularnewline
				\hline 
			\end{tabular}
			\par\end{centering}
		\caption{Redshift dependence of the pressure ratio.}
		
		\label{tab:redshifts}
	\end{table}
	\par\end{center}

It is, therefore, clear that $\rho_{\mathrm{c}}=10^{-26}\ \mathrm{g/cm^3}$ is a fairly good approximation for clouds in the CGM of Milky Way-like galaxies between redshifts 0 and 2. However, we also note that different works have different definitions
of the virial temperature, varying by a factor of $\approx2$. In addition,
the range of the CGM mass of the Milky Way proposed by \citet{Faerman2022} is quite wide, from $(3-10)\times10^{10}\ M_{\odot}$,
with $5\times10^{10}\ M_{\odot}$ only being a ``nominal'' value.

\end{document}